\documentclass[12pt,english]{iopart}
\usepackage{graphicx}
\usepackage{iopams}

\begin{document}

\title{Phase Control of Trapped Ion Quantum Gates}

\author{P. J. Lee, K.-A. Brickman, L. Deslauriers, P. C. Haljan, L.-M. Duan
and C. Monroe \\
}


\address{FOCUS Center and Department of Physics, University
of Michigan, Ann Arbor, MI 48109}

\begin{abstract}
There are several known schemes for entangling trapped ion quantum
bits for large-scale quantum computation. Most are based on an interaction
between the ions and external optical fields, coupling internal qubit
states of trapped-ions to their Coulomb-coupled motion. In this paper,
we examine the sensitivity of these motional gate schemes to phase
fluctuations introduced through noisy external control fields, and
suggest techniques to suppress the resulting phase decoherence.
\end{abstract}

\maketitle

Scalable quantum computing presents a direct application for the study
and control of large-scale quantum systems. The generally accepted
requirements for quantum hardware \cite{divincenzo00} include identifiable
two-level-systems for storing information in the form of quantum bits
(or qubits), and methods to externally manipulate and entangle qubits
through quantum logic gate operations. The implicit interconnects
represented by entangled quantum systems provide the power behind
quantum computation, giving rise to certain applications such as Shor's
factoring algorithm \cite{shor94} and Grover's search algorithm \cite{grover97a}
that exceed the capabilities of classical computers. However, in engineering
complex entangled states of many qubits, it is critical to control
the phase of the system of qubits and the phase of the classical control
parameters that guide the quantum gates. Techniques of quantum error-correction
\cite{shor95,steane96} appear essential to stabilize quantum computations,
but to reach fault-tolerant error-correction thresholds \cite{preskill99},
the host system must already possess a great deal of passive stability
and must be relatively insensitive to external noise.

One of the most promising quantum computing architectures is a system
of cold atomic ions confined in free space with electromagnetic fields
\cite{cirac-zoller95,wineland98b,kielpinski02}. Here, qubits are
stored in stable electronic states of each ion, and quantum information
can be processed and transferred through the collective quantized
motion of the Coulomb-coupled ion crystal. Applied electromagnetic
fields (usually from a laser) enable this coupling between internal
qubit states and external motional states, following several known
quantum gate schemes \cite{cirac-zoller95,sorensen99,Milburn00,Garcia-Ripoll03}.
Trapped ion quantum gates are thus highly susceptible to noise on
the applied laser fields in addition to ambient electric and magnetic
fields. In most cases, the relevant phases of the laser fields for
quantum gates can be sufficiently stable during the evolution of a
given gate, with gate speeds typically faster than about $100\mu s$.
However, for extended operations involving many successive gates,
it will be difficult to maintain optical phase stability over the
duration of the quantum computation.

From an engineering standpoint, the ability to perform gate operations
on any individual qubit or set of qubits with a given phase at any
step in a series of operations is requisite to a universal quantum
computer. We assume that all operations are synchronized to a local
oscillator in perfect resonance with the qubits. Each qubit initially
has an arbitrarily defined phase, and subsequent phase accumulations
from interactions must be tracked so that each operation account for
the phase of the individual qubit. These interactions are primarily
AC Stark shifts from the optical control fields \cite{Blatt03prl}
and Zeeman shifts from ambient magnetic fields. Our goal here is to
prescribe a set of gates that leave the qubits independent of the
optical phase of the driving field after each operation is complete
while enjoying passive isolation from Stark and Zeeman qubit phase
shifts.

In this paper, we consider several quantum gate schemes in the trapped
ion system, concentrating on a class of currently-favored quantum
gates that rely on a {}``spin-dependent force'' \cite{molmer99,sorensen99,Zagury99,Milburn00,leibfried03,Garcia-Ripoll03,Haljan05},
and are relatively insensitive to motional heating from noisy background
electric fields \cite{turchette00a,deslauriers04}. We will discuss
the sources of phase decoherence for these gates and describe methods
to suppress decoherence from slow phase drifts of the driving optical
fields. We find that certain gate schemes can be simultaneously insensitive
to background magnetic fields, making them quite robust for long-term
computations.

This paper is divided into three sections: Section 1 lays out the
background for the various quantum gate schemes for trapped ion system,
covering detailed steps for performing single qubit operations and
a discussion of the original Cirac-Zoller model with special attention
paid to the phases of the qubits \cite{cirac-zoller95}. Section 2
describes entangling gates using a spin-dependent force. We show how
special arrangements of the classical driving fields can suppress
slow phase decoherence from laser noise and external magnetic field
noise \cite{Haljan05}. Section 3 shows how to similarly suppress
long-term phase noise in a recent {}``fast'' gate scheme that relies
on impulsive spin-dependent optical forces \cite{Garcia-Ripoll03}.

\section{Background}

\begin{figure}
\begin{center}
\includegraphics[
  width=8cm,
  height=8cm,
  keepaspectratio,
  angle=-90]{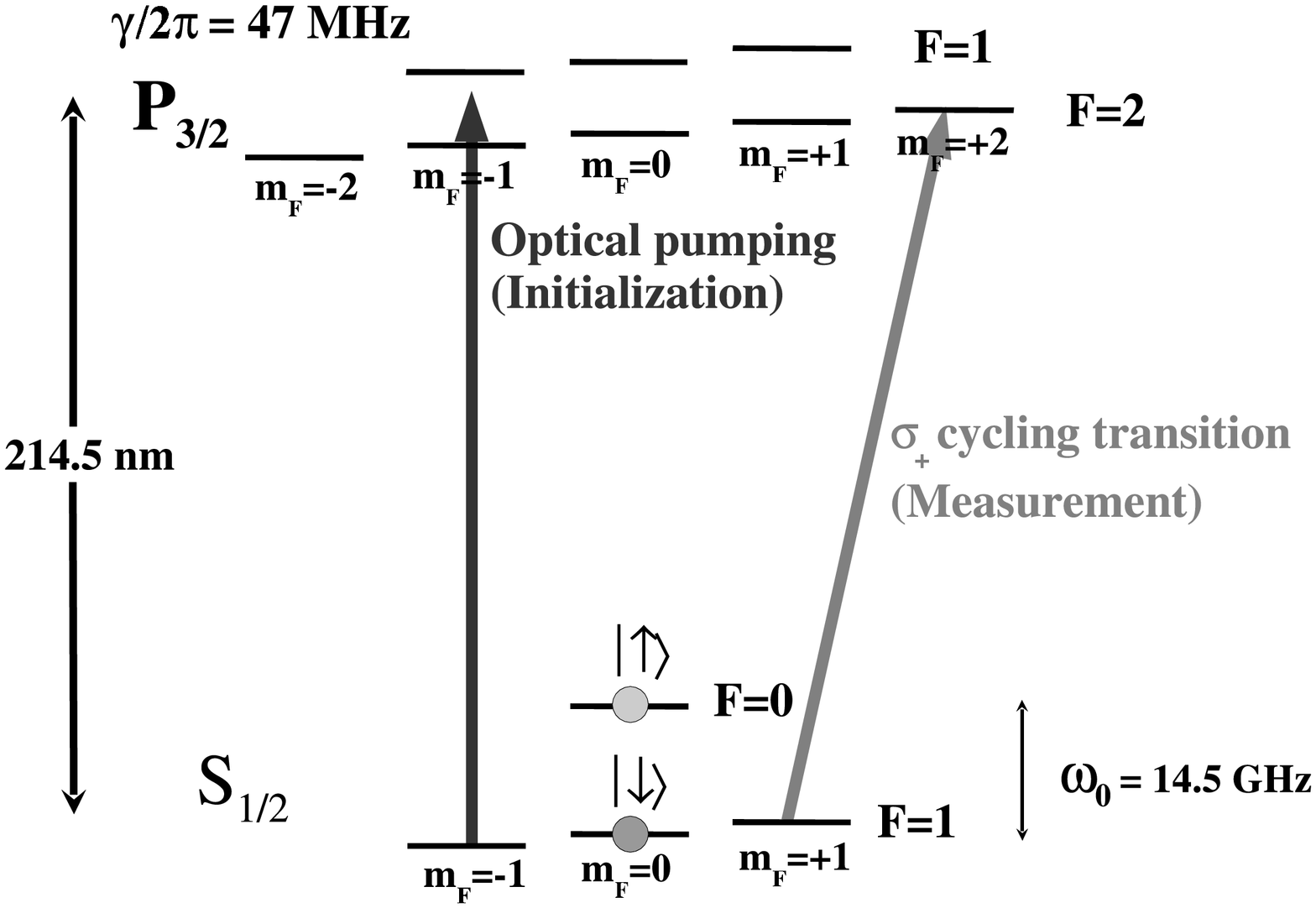}
\end{center}

\caption{$^{111}$Cd$^{+}$ as an example of a hyperfine qubit. The $^{2}S_{1/2}$
ground state electron configuration combined with the nuclear spin
$I=1/2$ creates an energy splitting of 14.5GHz due to the hyperfine
interaction. The states $\left|F=0,m_{F}=0\right\rangle $ and $\left|F=1,m_{F}=0\right\rangle $
are called {}``clock'' states at zero magnetic field since their
energy difference has no first order dependence on the magnetic field,
and are designated as the qubit states $\left|\uparrow\right\rangle $
and $\left|\downarrow\right\rangle $ respectively. A small external
magnetic field lifts the degeneracy of the $F=1$ states through the
Zeeman effect. Due to the large ground state hyperfine splitting,
the qubit can be initialized by optically pumping into the $\left|\uparrow\right\rangle =\left|F=0,m_{F}=0\right\rangle $
state. The qubit state can be measured by applying resonant $\sigma_{+}$
radiation to optically pump the $\left|\downarrow\right\rangle =\left|F=1,m_{F}=0\right\rangle $
state to the $\left|F=1,m_{F}=1\right\rangle $ state and drive a
cycling transition between the $^{2}S_{1/2}$, $F=1$ and the $^{2}P_{3/2}$,
$F=2$ excited state and collecting the resulting fluorescence. \label{cap:qubit}}
\end{figure}

Typical atomic ion species for quantum information applications such
as $^{9}$Be$^{+}$, $^{43}$Ca$^{+}$, and $^{111}$Cd$^{+}$ have
a single valence electron with a $^{2}S_{1/2}$ ground state and $^{2}P_{J}$
first excited electronic states. In isotopes with nonzero nuclear
spin, the $^{2}S_{1/2}$ ground states are split by the hyperfine interaction.
An applied static magnetic field $\mathbf{B}_{0}$ provides a quantization
axis and removes degeneracy in the ground state Zeeman levels. Two
states, one from each ground state hyperfine level, are designated
as the qubit states, denoted by $\uparrow_{i}$ and $\downarrow_{i}$
for each ion $i$ and separated by energy $E_{\uparrow}-E_{\downarrow}=\hbar\omega_{0}$.
At certain values of $\mathbf{B}_{0}$, this energy splitting can be insensitive
to magnetic field fluctuations to first order, forming a qubit that
can have particularly good phase stability. Such qubit levels are
termed {}``clock states'' because their stability is exploited in
atomic clocks \cite{bollinger91}. The qubit frequency splitting $\omega_{0}$
is usually in the microwave range and large compared with the radiative
line-width $\gamma$ of excited electronic states, therefore the qubit
can be measured with high fidelity by resonant pumping to a cycling
transition between one hyperfine state and an excited electronic state
\cite{wineland98b}. Initialization of the qubits can also be accomplished
with high accuracy using optical pumping techniques (Figure \ref{cap:qubit}).

We assume ions are confined in a linear Paul trap \cite{Paul90} with
a combination of static and radio frequency (rf) electric quadrupole
potentials \cite{dehmelt67}. When the ions are sufficiently cold,
their Coulomb repulsion balances the external confinement forces,
and the ions form stationary crystals with the residual motion described
by coupled harmonic oscillatory motion in the trap. The number of
collective modes scales linearly with the number of atoms $N$ in
the trap, making it difficult to isolate and control all modes of
oscillation for large numbers of ions. To circumvent this difficulty,
an architecture has been proposed for shuttling ions between multiple
trap regions in a trap structure such that it is only necessary to
localize a small number of ions at a given time \cite{kielpinski02}.
Two-qubit quantum gates in addition to arbitrary single qubit rotations
are sufficient for the engineering of arbitrarily complex entangled
states \cite{divincenzo00}, so we focus on the operation of quantum
gates on $N=2$ ions, although extensions to larger numbers of ions
is straightforward.

The ions arrange themselves along the weakest ($x$) axis of the confinement
potential, and the position operator of each ion can be written as
$\hat{\mathbf{R}}_{i}=\mathbf{R}_{0,i}+\hat{\mathbf{r}}_{i}$, with
the operator $\hat{\mathbf{r}}_{i}$ describing the small quantum
harmonic oscillations of each ion about its equilibrium position $\mathbf{R}_{0,i}$.
Of the six normal modes of oscillation for the two ions, only the
two axial normal modes will be considered for simplicity (see section
1.1): the center-of-mass coordinate $\hat{q}_{1}=(\hat{x}_{1}+\hat{x}_{2})/\sqrt{2}$
and a {}``stretch'' coordinate $\hat{q}_{2}=(\hat{x}_{1}-\hat{x}_{2})/\sqrt{2}$,
where $\hat{x}_{i}$ is the component of $\hat{\mathbf{r}}_{i}$ along
the $x$-axis. The base Hamiltonian for the collective system is \begin{equation}
\hat{H}_{0}=\sum_{i=1,2}\hbar\omega_{0}\left|\uparrow_{i}\right\rangle \left\langle \uparrow_{i}\right|+\sum_{\nu=1,2}\hbar\omega_{\nu}\hat{a}_{\nu}^{\dagger}\hat{a}_{\nu}\label{eq: unperturbed Hamiltonian}\end{equation}
 where $\omega_{0}$ is the frequency difference between the two qubit
states; $\omega_{1}$ and $\omega_{2}=\sqrt{3}\omega_{1}$ are the
frequencies associated with the center-of-mass and stretch modes,
respectively; and $\hat{a}_{\nu}^{\dagger}$ and $\hat{a}_{\nu}$
are their respective harmonic oscillator creation and annihilation
operators. The first term in the Hamiltonian describes the internal
energy of the ions, and the second term describes the external vibrational
energy of the system.

Single qubit rotations between hyperfine qubit levels (not involving
ion motion) can be performed by applying appropriate radiation fields.
For example, a resonant microwave field can directly couple the qubit
levels through a magnetic dipole interaction, resulting in coherent
Rabi oscillations between the two qubit states. Alternatively, optical
stimulated-Raman transitions can be employed \cite{monroe95b}, using
two optical sources that coherently couple the qubit states through
excited $^{2}P_{J}$ electronic states, as discussed next.

\subsection{Coherent Interaction between Trapped-Ion Hyperfine Qubits and Optical
Fields}

\begin{figure}
\begin{center}
\includegraphics[
  width=8cm,
  height=5cm,
  keepaspectratio,
  angle=-90]{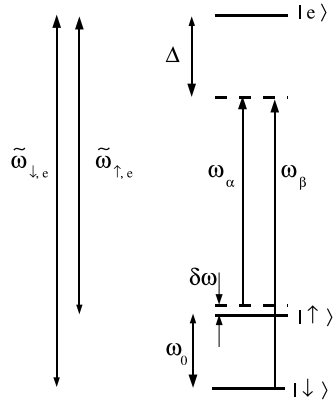}
\end{center}

\caption{Qubit with coherent Raman coupling via an excited state. Fields at
frequency $\omega_{\alpha}$ and $\omega_{\beta}$ couple the qubit
levels $\left|\uparrow\right\rangle $ and $\left|\downarrow\right\rangle $
via the excited state $\left|e\right\rangle $. The fields are detuned
from the excited state resonances $\tilde{\omega}_{\uparrow,e}$ and
$\tilde{\omega}_{\downarrow,e}$ by frequency $\Delta$. \label{cap:two_level}}
\end{figure}

An optical coupling between the hyperfine qubit states and an excited
electronic state of each ion can be exploited to entangle qubit states
with collective motional states, forming the backbone of most trapped
ion quantum logic gates. As shown in Figure \ref{cap:two_level},
we assume each of two trapped ions consists of three levels: the two
ground state qubit levels $\left|\uparrow_{i}\right\rangle $ and
$\left|\downarrow_{i}\right\rangle $ and an excited electronic state
$\left|e_{i}\right\rangle $ having respective optical frequency spans
$\tilde{\omega}_{\uparrow,e}$ and $\tilde{\omega}_{\downarrow,e}=\tilde{\omega}_{\uparrow,e}+\omega_{0}$.
Two optical fields $\mathbf{E}_{l}(\mathbf{r})=\tilde{E}_{l}(\mathbf{r})\cos(\mathbf{k}_{l}\mbox{$\cdot$}\mathbf{r}-\omega_{l}t-\phi_{l})\mbox{\boldmath$\epsilon$}_{l}$
with $l=\alpha,\beta$ and polarization $\mbox{\boldmath$\epsilon$}_{l}$, connect each of the qubit levels $\left|\uparrow_{i}\right\rangle $
and $\left|\downarrow_{i}\right\rangle $, respectively, to state
$\left|e_{i}\right\rangle $ through electric dipole operators $\mbox{\boldmath$\mu$}_{\uparrow}$
and $\mbox{\boldmath$\mu$}_{\downarrow}$. We assume that the optical fields
have a difference frequency $\omega_{\beta}-\omega_{\alpha}=\omega_{0}+\delta\omega$
and are both detuned from the excited-state resonance by $\Delta=\tilde{\omega}_{\uparrow,e}-\omega_{\alpha}$,
as drawn in Figure \ref{cap:two_level}. These fields evaluated at
the ion's position $\mathbf{E}_{l}(\hat{\mathbf{R}}_{i})$ are what
ultimately couples the spin to the motion.

The interaction can be transformed to a rotating frame at frequency
$\omega_{\alpha}$ in order to remove all terms varying with optical
frequencies, and under the usual optical rotating wave approximation
(RWA), the interaction Hamiltonian between the fields and the ions
is

\begin{eqnarray}
\fl \hat{H}_{I}=\frac{\hbar}{2}\sum_{i=1,2}\bigg[-\left(g_{\uparrow,\alpha,i}e^{i\mathbf{k}_{\alpha}\cdot\hat{\mathbf{R}}_{i}-i\phi_{\alpha}}\left|e_{i}\right\rangle \left\langle \downarrow_{i}\right|+h.c.\right)-\left(g_{\downarrow,\beta,i}e^{i\mathbf{k}_{\beta}\cdot\hat{\mathbf{R}}_{i}-i\phi_{\beta}}e^{-i(\delta\omega)t}\left|e_{i}\right\rangle \left\langle \uparrow_{i}\right|+h.c.\right)
\nonumber \\
\lo +\Delta\left|e_{i}\right\rangle \left\langle e_{i}\right|\bigg]\label{General_H}
\end{eqnarray}
 In this expression, the dipole coupling strengths between qubit state
$\left|m_{i}\right\rangle =\left|\uparrow_{i}\right\rangle ,\left|\downarrow_{i}\right\rangle $
and excited state $\left|e_{i}\right\rangle $ from laser field $l$
on ion $i$ are given by $\hbar g_{m,l,i}=-$$\mbox{\boldmath$\mu$}_{m}\cdot\mbox{\boldmath$\epsilon$}_{l}\tilde{E}_{l}(\mathbf{R}_{0,i})$$/2$.

For most of the remainder of this paper (outside of Section 3), we
assume that the detuning $\Delta$ of the optical fields from electronic
resonance is much larger than the excited state line-width $\gamma$
and the couplings $\left|g_{m,l,i}\right|^{2}$, so that spontaneous
emission during the optical coupling is negligible \cite{wineland98b}
and the excited state $\left|e_{i}\right\rangle $ can be adiabatically
eliminated. Applying RWA on the microwave frequencies, we find

\begin{eqnarray}
\fl \hat{H}_{I}=-\frac{\hbar}{2}\sum_{i=1,2}\bigg[\left(\Omega_{i}e^{-i(\Delta\mathbf{k}\cdot\mathbf{\hat{R}}_{i}-(\delta\omega)t-\Delta\phi)}\left|\uparrow_{i}\right\rangle \left\langle \downarrow_{i}\right|+h.c.\right)
\nonumber\\
\lo +\chi_{\downarrow,i}\left|\downarrow_{i}\right\rangle \left\langle \downarrow_{i}\right|+\chi_{\uparrow,i}\left|\uparrow_{i}\right\rangle \left\langle \uparrow_{i}\right|\bigg]\label{AdElim_H}\end{eqnarray}
 where $\Delta\mathbf{k}=\mathbf{k}_{\beta}-\mathbf{k}_{\alpha}$
and $\Delta\phi=\phi_{\beta}-\phi_{\alpha}$ are the differences in
the wave-vector and the phase of the two applied fields, $\Omega_{i}=g_{\uparrow,\alpha,i}g_{\downarrow,\beta,i}^{*}/2\Delta$
is the {}``base Rabi frequency'' directly coupling the qubit states
of ion $i$, and $\chi_{m,i}=\left(\left|g_{m,\alpha,i}\right|^{2}+\left|g_{m,\beta,i}\right|^{2}\right)/2$
corresponds to the AC Stark shifts of the qubit level $\left|m_{i}\right\rangle $
of ion $i$ by both optical fields.

For simplicity, we assume $\Delta\mathbf{k}$ is parallel to the x-axis
($\Delta k=\left|\Delta\mathbf{k}\right|$), so that the interaction
deals only with axial motion (although it is straightforward to treat
the more general case). We substitute the x-component of the position
operator $\hat{\mathbf{R}}_{i}$ for ion $i$ by
\begin{equation}
\hat{X}_{i}=X_{0,i}+\frac{q_{1}}{\sqrt{2}}(\hat{a}_{1}+\hat{a}_{1}^{\dagger})\pm\frac{q_{2}}{\sqrt{2}}(\hat{a}_{2}+\hat{a}_{2}^{\dagger})\label{eq: normal modes}\end{equation}
 where $q_{\nu}=\sqrt{\hbar/(2M\omega_{\nu})}$ is the root mean square
spatial spread of the ground state wavepacket for normal mode $\nu$
of oscillation in the trap, $M$ is the single ion mass, and the plus
(minus) sign refers to ion $i=1$ ($i=2)$. In the interaction frame
of the vibrational levels, Equation \ref{AdElim_H} becomes
\begin{eqnarray}
\fl \hat{H}_{I}=-\frac{\hbar}{2}\sum_{i=1,2}\bigg[(\Omega_{i}e^{-i\left[\eta_{1}(\hat{a}_{1}e^{-i\omega_{1}t}+\hat{a}_{1}^{\dagger}e^{i\omega_{1}t})\pm\eta_{2}(\hat{a}_{2}e^{-i\omega_{2}t}+\hat{a}_{2}^{\dagger}e^{i\omega_{2}t})\right]}e^{i(\delta\omega)t}e^{-i(\Delta kX_{0,i}-\Delta\phi)}\left|\uparrow_{i}\right\rangle \left\langle \downarrow_{i}\right|+h.c.)\nonumber \\
\lo +\chi_{\beta,i}\left|\downarrow_{i}\right\rangle \left\langle \downarrow_{i}\right|+\chi_{\alpha,i}\left|\uparrow_{i}\right\rangle \left\langle \uparrow_{i}\right|\bigg]
\end{eqnarray}
 The Lamb-Dicke parameters are defined by $\eta_{1}=\Delta kq_{1}/\sqrt{2}$
and $\eta_{2}=\Delta kq_{2}/\sqrt{2}=\eta_{1}/\sqrt[4]{3}$, representing
the coupling strength between the fields and each normal mode.

The above treatment can be generalized to the case of multiple optical
sources that connect both qubit states to any number of excited states,
resulting in higher order expressions for $\Omega_{i}$ and $\chi_{m,i}$.
Here however, we are mainly interested in the sensitivity of entangling
gate operations on the optical phases $\phi_{l}$. The net optical
phase appearing in the coupling Hamiltonian (Equation \ref{AdElim_H})
is sensitive only to the phase \textit{difference} $\Delta\phi=\phi_{\beta}-\phi_{\alpha}$
between the two optical fields, so that when both fields are generated
from a single laser and modulator, fluctuations in the optical phase
of the laser source become common mode and do not lead to decoherence
\cite{thomas82}. However, in order to couple the qubits with the
motion for entangling quantum gates, the optical sources are generally
non-copropagating ($\Delta k$$\neq0$), opening up the sensitivity
to decoherence from fluctuations in relative beam path lengths or
ions' positions through the phase factor $e^{i\Delta kX_{0,i}-i\Delta\phi}$.
This requires interferometric stability between the optical paths
of the fields $E_{\alpha}$ and $E_{\beta}$, which should be feasible
over short times using stable optical mounts and indexing the laser
beams to the trap structure itself. However, over the long time scale
represented by an extended quantum computation, drifts in the phase
$e^{i\Delta kX_{0,i}-i\Delta\phi}$ can be a serious source of decoherence.

\subsection{Resolved-Sideband Limit}

Equation \ref{AdElim_H} includes direct couplings between qubit states,
and entangling couplings between qubit states and trapped ion motional
states. We consider the case where the base Rabi frequencies $\Omega_{i}$
are much smaller than the vibrational frequencies $\omega_{\nu}$
of the ions in the trap. In this case, the difference frequency $\delta\omega$
of the optical sources can be tuned to particular values so that a
single stationary term emerges from the above Hamiltonian, and all
others couplings can be neglected under the rotating wave approximation.
In this regime, as seen from the rest frame of the ions, the applied
laser fields acquire resolved frequency-modulation sidebands from
the ions' harmonic vibration. We concentrate on three spectral features:
the {}``carrier'', the first upper and lower sidebands, each selected
by appropriate tuning of the radiation field difference frequency.

\subsubsection{The carrier}

\begin{figure}
\begin{center}
\includegraphics[
  width=8cm,
  height=8cm,
  keepaspectratio,
  angle=-90]{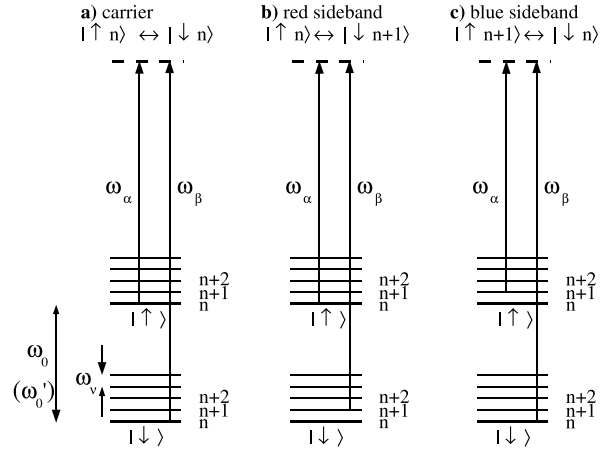}
\end{center}

\caption{Stimulated Raman transition between vibrational levels. Coupling
depends on the beat note of the two Raman fields $\omega_{\beta}-\omega_{\alpha}$:
a) $\omega_{0}'$ for carrier transition, b) $\omega_{0}'-\omega_{\nu}$
for first red sideband transition, and c) $\omega_{0}'+\omega_{\nu}$
for first blue sideband transition. The qubit frequency splitting
shifts from $\omega_{0}$ to $\omega_{0}'$ due to AC Stark effect
when the fields are turned on. \label{cap:Stimulated-Raman-transition}}
\end{figure}

When the difference frequency between the optical sources is tuned
to the free-ion qubit resonance (compensating for possible differential
Stark shifts, assumed to be equal for the two ions), then $\omega_{\beta}-\omega_{\alpha}=\omega_{0}+\chi_{\uparrow,i}-\chi_{\downarrow,i}\equiv \omega_{0}'$
(Figure \ref{cap:Stimulated-Raman-transition}a ), and we find the
stationary term in Equation \ref{AdElim_H} is given by \cite{wineland98b}
\begin{equation}
\hat{H}_{I}^{car}=-\frac{\hbar}{2}\sum_{i=1,2}\sum_{n_{1},n_{2}}\left(\Omega_{i}\mathcal{D}_{n_{1},n_{2}}e^{i(\Delta kX_{0,i}-\Delta\phi)}\hat{\sigma}_{+}^{(i)}+h.c.\right)\left|n_{1},n_{2}\right\rangle \left\langle n_{1},n_{2}\right|.\label{carrier}\end{equation}
 This {}``carrier'' interaction describes simple Rabi flopping between
the qubit states in each ion, where the qubit raising and lowering
operators are defined by $\hat{\sigma}_{+}^{(i)}=\left|\uparrow_{i}\right\rangle \left\langle \downarrow_{i}\right|$
and $\hat{\sigma}_{-}^{(i)}=(\hat{\sigma}_{+}^{(i)})^{\dagger}=\left|\downarrow_{i}\right\rangle \left\langle \uparrow_{i}\right|$.
Also, $\mathcal{D}_{n_{1},n_{2}}=e^{-\frac{1}{2}(\eta_{1}^{2}+\eta_{2}^{2})}\mathcal{L}_{n_{1}}(\eta_{1}^{2})\mathcal{L}_{n_{2}}(\eta_{2}^{2})$
is the Debye-Waller factor that exponentially suppresses the carrier
coupling due to ion motion described by vibrational quantum numbers
$n_{\nu}$ in each mode $\nu$ of motion, with $\mathcal{L}_{n_{\nu}}(z)$
being a Laguerre polynomial of order $n_{\nu}$ \cite{wineland98b}.
When the ions are confined to the Lamb-Dicke limit (LDL) where $\eta_{\nu}^{2}(n_{\nu}+1/2)\ll1$,
then $\mathcal{D}_{n_{1},n_{2}}\simeq1$. However, if the motion of
either mode is not in a pure eigenstate of harmonic motion and outside
of the LDL, then the Rabi frequency will depend upon the noisy motional
quantum state and lead to qubit decoherence. In order to avoid this
problem, carrier operations are often performed with copropagating
Raman beams ($\eta_{\nu}=0$) thereby forcing $\mathcal{D}_{n_{1},n_{2}}=1$.
A carrier transition using a copropagating geometry can also be insensitive to the phase noise of the source laser, making it ideal for single
qubit rotations.

\subsubsection{The first lower (red) sideband}

When the difference frequency between the optical sources is tuned
lower than the free-atom qubit resonance by the vibrational frequency
$\omega_{\nu}$ of mode $\nu$ (again compensating for differential
Stark shifts), $\omega_{\beta}-\omega_{\alpha}=\omega_{0}'-\omega_{\nu}$
(Figure \ref{cap:Stimulated-Raman-transition}b) and we find the
stationary term in Equation \ref{AdElim_H} is \cite{wineland98b}
\begin{equation}
\mathcal{\hat{H}}_{I}^{rsb}=-\frac{\hbar}{2}\sum_{i=1,2}\left(\eta_{\nu}\Omega_{i}e^{i(\Delta kX_{0,i}-\Delta\phi)}\mathcal{D}_{n_{\nu},n_{\nu'}}^{\prime}\hat{\sigma}_{+}^{(i)}\hat{a}_{\nu}+h.c.\right).\label{rsb}\end{equation}
 This {}``red sideband{}`` interaction describes Rabi flopping between
the coupled qubit-motional states $\left|\downarrow,n_{\nu}\right\rangle $
and $\left|\uparrow,n_{\nu}-1\right\rangle $ in each ion, where $\mathcal{D}_{n_{\nu},n_{\nu'}}^{\prime}=e^{-\frac{1}{2}(\eta_{1}^{2}+\eta_{2}^{2})}\frac{\mathcal{L}_{n_{\nu}-1}^{1}(\eta_{\nu}^{2})}{n_{\nu-1}!}\mathcal{L}_{n_{\nu'}}(\eta_{\nu'}^{2})$
is the Debye-Waller factor for the first sideband, with $\nu'\neq\nu$
the {}``spectator'' mode of motion. Here, $\mathcal{L}_{n}^{1}(z)$
is an associated Laguerre polynomial. This interaction is the Jaynes-Cummings
Hamiltonian \cite{jaynes63} where energy is exchanged between the
internal qubit and the external harmonic oscillator states.

\subsubsection{The first upper (blue) sideband}

When the difference frequency between the optical sources is tuned
higher than the free-atom qubit resonance by the vibrational frequency
$\omega_{\nu}$ of mode $\nu$ (once again compensating for differential
Stark shifts), $\omega_{\beta}-\omega_{\alpha}=\omega_{0}'+\omega_{\nu}$
(Figure \ref{cap:Stimulated-Raman-transition}c) and we find the
stationary term in Equation \ref{AdElim_H} is now \cite{wineland98b}\begin{equation}
\mathcal{\hat{H}}_{I}^{bsb}=-\frac{\hbar}{2}\sum_{i=1,2}\left(\eta_{\nu}\Omega_{i}e^{i(\Delta kX_{0,i}-\Delta\phi)}\mathcal{D}_{n_{\nu},n_{\nu'}}^{\prime}\hat{\sigma}_{+}^{(i)}\hat{a}_{\nu}^{\dagger}+h.c.\right).\label{bsb}\end{equation}
 This {}``blue sideband'' or anti-Jaynes-Cummings interaction describes
Rabi flopping between the coupled qubit-motional states $\left|\downarrow,n_{\nu}-1\right\rangle $
and $\left|\uparrow,n_{\nu}\right\rangle $ in each ion.

\subsection{The Cirac-Zoller Gate}

The original Cirac-Zoller (CZ) scheme \cite{cirac-zoller95,monroe95b,Blatt03}
illustrates how entanglement between trapped ion qubits can be achieved
through coupling of each qubit to a common mode of motion in the trap.
The CZ scheme allows the operation of a controlled-NOT gate between
two trapped ion qubits, flipping the state of a target qubit (e.g.,
$\left|\downarrow_{2}\right\rangle \leftrightarrow\left|\uparrow_{2}\right\rangle $)
only when the control qubit is, say, in state $\left|\downarrow_{1}\right\rangle $.
This can be accomplished by cooling a collective motional mode $\nu$
of the two ions to the $\left|0_{\nu}\right\rangle $ ground state
and performing the following three steps:

(1) A carrier $\pi/2$-pulse on the target qubit with associated phase
$\phi$,

(2) A $\pi$ phase gate on two ions

(3) A carrier $-\pi/2$-pulse on the target qubit with phase $\phi$
(step (i) reversed).

\begin{figure}
\begin{center}
\includegraphics[
  width=8cm,
  height=8cm,
  keepaspectratio,
  angle=-90]{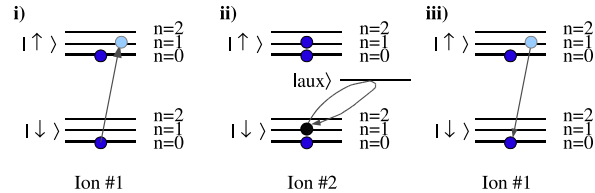}
\end{center}

\caption{Cirac-Zoller Gate Scheme \cite{cirac-zoller95}. A phase gate is
constructed by performing the sequence illustrated here: i) a $\pi$-pulse
on the first blue sideband on the first ion to map the internal state
to the collective vibrational state; ii) a $2\pi$-pulse between the
$\left|\downarrow,n=1\right\rangle $ state and an auxiliary state
$\left|aux\right\rangle $ on the second ion, resulting in a $\pi$
phase shift on the state $\left|\downarrow,n=1\right\rangle $; iii)
a $\pi$-pulse on the first blue sideband on the first ion to map
the vibrational state back to the internal state. A controlled-NOT
gate can be constructed from a phase gate with a $\pi/2$-pulse on
the second qubit before and after the phase gate. While the phase
gate by definition maintains strict control of the qubit phase, the
controlled-NOT gate relies on the two additional $\pi/2$-pulses having
a particular phase with respect to the qubit. \label{cap:Cirac-Zoller}}
\end{figure}

Steps (1) and (3) are simply carrier couplings on the target qubit
ion, achieved by focusing radiation on the target ion only ($\Omega_{1}=0$)
and applying the radiation for a time $t_{\pi/2}$ ($|\Omega_{2}|t_{\pi/2}=\pi/2$).
Step (1) results in the evolution \begin{equation}
\alpha\left|\uparrow_{2}\right\rangle +\beta\left|\downarrow_{2}\right\rangle \rightarrow\frac{\left(\alpha+e^{-i\phi}\beta\right)}{\sqrt{2}}\left|\uparrow_{2}\right\rangle +\frac{\left(\beta-e^{i\phi}\alpha\right)}{\sqrt{2}}\left|\downarrow_{2}\right\rangle .\label{pi/2 carrier}\end{equation}
 Step (3) is identical to step (1) except the phase is shifted by
$\pi.$ There are many ways to implement step (2), one of which is:
i) a $\pi$-pulse blue sideband that maps the internal qubit state
of the control qubit to the collective state of the ion pair, ii)
a $2\pi$-pulse coupling the $\left|\downarrow_{2}\right\rangle \left|n=1\right\rangle $
state exclusively to an auxiliary level, and iii) a $\pi$-pulse on
the blue sideband to map the collective motional state back to the
control bit (see Figure \ref{cap:Cirac-Zoller}). The net effect of
these steps produces the following phase gate: \begin{equation}
\begin{array}{rrr}
\left|\uparrow\uparrow\right\rangle  & \rightarrow & \left|\uparrow\uparrow\right\rangle \\
\left|\uparrow\downarrow\right\rangle  & \rightarrow & \left|\uparrow\downarrow\right\rangle \\
\left|\downarrow\uparrow\right\rangle  & \rightarrow & \left|\downarrow\uparrow\right\rangle \\
\left|\downarrow\downarrow\right\rangle  & \rightarrow & -\left|\downarrow\downarrow\right\rangle \end{array}.\label{eq: phase gate}\end{equation}
Here every state maintains a constant amplitude and the phase is well-defined.
However, steps (1) and (3) contribute an additional phase to the controlled-NOT
gate:\begin{equation}
\begin{array}{rrr}
\left|\uparrow\uparrow\right\rangle  & \rightarrow & \left|\uparrow\uparrow\right\rangle \\
\left|\uparrow\downarrow\right\rangle  & \rightarrow & \left|\uparrow\downarrow\right\rangle \\
\left|\downarrow\uparrow\right\rangle  & \rightarrow & e^{i\phi}\left|\downarrow\downarrow\right\rangle \\
\left|\downarrow\downarrow\right\rangle  & \rightarrow & e^{i\phi}\left|\downarrow\uparrow\right\rangle \end{array}.\label{eq: c-NOT gate}\end{equation}
The phase of the Cirac-Zoller controlled-NOT gate therefore depends
solely on the phase of the $\pi/2$-pulse single qubit rotations.
As mentioned in section 1.1, the sensitivity of single qubit rotations
to optical phase can be removed using copropagating Raman beams requiring
only a stable microwave source driving an optical modulator. This
conversion between a phase gate and a controlled-NOT gate is extremely
useful for many entangling gate schemes, since phase gates are more
intuitive to construct and have an inherently well-defined phase.

The CZ model for trapped ion quantum logic gates has many drawbacks,
including the need for individually addressing the ions with optical
sources, and the requirement that the motion be prepared in a pure
state of collective motion, usually through laser-cooling to the $\left|0_{\nu}\right\rangle $
state. In the remainder of this paper, we consider improved schemes
for trapped ion quantum gates that do not have these requirements.
In some cases, we will see that the sensitivity to the optical phase
$e^{i\Delta kX_{0,i}-i\Delta\phi}$ can also be suppressed.

\section{Spin-Dependent Forces in the Resolved-Sideband Limit}

Unlike the Cirac-Zoller gate where the internal qubit state of one
ion is directly transferred to particular eigenstates of motion, entangling
gates using spin-dependent forces coherently displace the initial
motional state in the position/momentum phase space, a process through
which each spin state can acquire an independent geometric phase\cite{Milburn00,molmer99,sorensen99,leibfried03,Garcia-Ripoll03}.
The nonlinearity in these phases can result in a final state that
can no longer be separated into two independent qubit subspaces, thus
entangling the internal states of the two ions. This produces a phase
gate similar to the Cirac-Zoller scheme, which can then be converted
to a controlled-NOT gate when combined with single qubit rotations.

\begin{figure}
\begin{center}
\includegraphics[
  width=8cm,
  height=6cm,
  keepaspectratio]{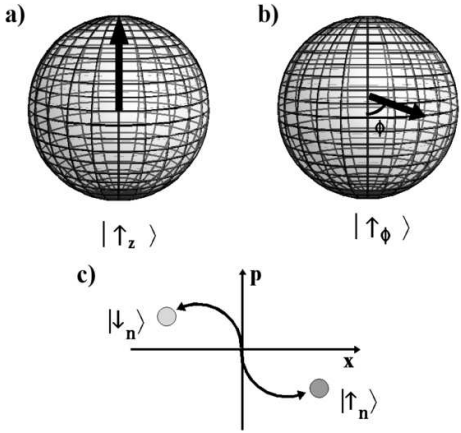}
\end{center}

\caption{Representation of the eigenstate of a) $\mbox{\boldmath$\sigma$}\cdot\mathbf{z}$
and b) $\mbox{\boldmath$\sigma$}\cdot\mbox{\boldmath$\phi$}$ on the Bloch sphere corresponding
to the eigenvalue $+1$. A spin-dependent force creates two separate
coherent states in phase space corresponding to the eigenstates of
$\mbox{\boldmath$\sigma$}\cdot\mathbf{n}$, as represented in c), thus entangling
the internal spin with the external motion of the ion. \label{cap: Bloch spheres}}
\end{figure}

In this section, we focus on gates in the resolved-sideband limit
where the interaction time is much longer than the trap period. The
interaction Hamiltonian is proportional to $\hat{\mbox{\boldmath$\sigma$}}_{1}\cdot\mathbf{n}_{1}\otimes\hat{\mbox{\boldmath$\sigma$}}_{2}\cdot\mathbf{n}_{2}$,
where $\hat{\mbox{\boldmath$\sigma$}}_{i}$ is the Pauli spin matrix operating
on the internal qubit states, and $\mathbf{n}_{i}$ is a unit vector
pointing in a particular direction on the Bloch sphere for ion $i$.
The eigenstates of $\hat{\mbox{\boldmath$\sigma$}}_{1}\cdot\mathbf{n}_{1}\otimes\hat{\mbox{\boldmath$\sigma$}}_{2}\cdot\mathbf{n}_{2}$ each experiences a different
force from the interaction (see Figure \ref{cap: Bloch spheres}).
The gates are categorized according to the direction of $\mathbf{n}_{i}$:
in a {}``$\hat{\sigma}_{z}$ gate'', the differential force is
applied via a differential AC Stark shift on the states $\left|\uparrow_{i}\right\rangle $
and $\left|\downarrow_{i}\right\rangle $ induced by the laser fields
\cite{leibfried03}. However, clock states exhibit no differential
AC Stark shift when the Raman detuning $\Delta$ is large compared
to the qubit frequency splitting $\omega_{0}$ (see Appendix A), so
the only available qubit states for a $\hat{\sigma}_{z}$ gate
are thus susceptible to magnetic field fluctuations. In a {}``$\hat{\sigma}_{\phi}$
gate'', optical fields driving spin flips and coupling to the motion
produces a differential force between eigenstates of $\hat{\mbox{\boldmath$\sigma$}}_{i}\cdot\mbox{\boldmath$\phi$}_{i}$,
where the unit vector $\mbox{\boldmath$\phi$}_{i}=\cos(\phi_{i})\mathbf{x}+\sin(\phi_{i})\mathbf{y}$
lies on the equatorial plane of the Bloch sphere\cite{molmer99,sorensen99,sackett00,Haljan05}.
Although this gate is compatible with clock states, the optical beam
configuration can give rise to extreme sensitivity of the qubit phase
on the optical phase of the driving field, which can be the limiting
factor in the fidelity of the gate\cite{sackett00}. In this section,
we propose a method to cancel this phase dependence on the optical
field, relaxing the constraint on long-term interferometric stability
between the two Raman beam paths for the entirety of a multi-gate
sequence quantum algorithm.

\subsection{Forced Quantum Harmonic Oscillator}

In order to understand the spin-dependent force, we start by considering
the effects when a force is applied to a harmonic oscillator. In general,
a forced harmonic oscillator has a Hamiltonian of the form \cite{merzbacher}
\begin{equation}
\hat{H}=\hbar\omega(\hat{a}^{\dagger}\hat{a}+\frac{1}{2})+f^{*}(t)x_{0}\hat{a}+f(t)x_{0}\hat{a}^{\dagger},\label{eq: forced HO}\end{equation}
where $\hat{a}$ and $\hat{a}^{\dagger}$ are the annihilation and
creation operators respectively, and $x_{0}=\sqrt{\hbar/(2M\omega)}$
is the root mean square spatial spread of the ground state wavepacket.
The first term is the unperturbed Hamiltonian for the harmonic oscillator of frequency $\omega$,
and the last two terms correspond to an external time-dependent force
$f(t)$ applied to the system. In the interaction picture \begin{equation}
\hat{H}_{I}(t)=f^{*}(t)x_{0}\hat{a}e^{-i\omega t}+f(t)x_{0}\hat{a}^{\dagger}e^{i\omega t}.\label{eq: interaction picture 1}\end{equation}
Assuming the force $f(t)=Fe^{-i(\omega-\delta)t}/2$ is detuned from resonance by frequency $\delta\ll\omega$, then the interaction Hamiltonian can be rewritten
as\begin{equation}
\hat{H}_{I}(t)=\frac{F^{*}x_{0}}{2}\hat{a}e^{-i\delta t}+\frac{Fx_{0}}{2}\hat{a}^{\dagger}e^{i\delta t}.\label{eq: Interaction picture 2}\end{equation}
The state after an interaction time $t$ is prescribed by the time-evolution
operator \begin{equation}
\hat{U}(t)=\exp\left\{ -\frac{i}{\hbar}\left(\int_{0}^{t}\hat{H}_{I}(t')dt'+\frac{1}{2}\int_{0}^{t}dt'\int_{0}^{t'}dt''[\hat{H}_{I}(t'),\hat{H}_{I}(t'')]+...\right)\right\} .\label{eq: evolution operator}\end{equation}

\begin{figure}
\begin{center}
\includegraphics[
  width=8cm,
  height=8cm,
  keepaspectratio,
  angle=-90]{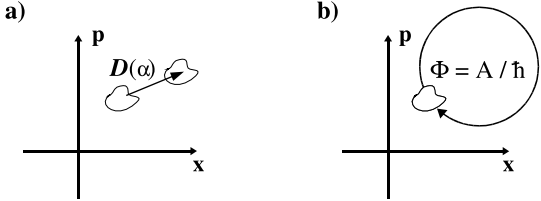}
\end{center}

\caption{a) Displacement in phase space, in a frame rotating at the natural frequency of the harmonic oscillator. The displacement operator translates
motional states in position/momentum phase space without distortion.
b) For a force detuned from resonance, the motional state follows a circular path.  For a closed trajectory, the quantum state acquires a geometric phase $\phi=A/\hbar$ in a
round-trip orbit, where $A$ is the area enclosed by the trajectory.
\label{cap:Displacement}}
\end{figure}

If we consider only the first term in the exponent of the evolution
operator and substituting in the interaction Hamiltonian from Equation
\ref{eq: Interaction picture 2}, the resulting operator is exactly
the displacement operator \begin{equation}
\hat{D}(\alpha)=e^{\alpha\hat{a}^{\dagger}+\alpha^{*}\hat{a}},\label{eq: Displacement operator}\end{equation}
with $\alpha$ defined as \begin{equation}
\alpha(t)=-\frac{i}{\hbar}\int_{0}^{t}\frac{Fx_{0}}{2}e^{i\delta t'}dt'.\label{eq: alpha general}\end{equation}
The displacement operator translates motional states in position/momentum
phase space without distortion (Figure \ref{cap:Displacement}). For
example, a displacement on an initial ground state of motion results
in a coherent state $\left|\alpha\right\rangle =\hat{D}(\alpha)\left|0\right\rangle $,
where the final state is defined in terms of number states as $\left|\alpha\right\rangle =e^{-\frac{1}{2}\left|\alpha\right|^{2}}\sum_{n=0}^{\infty}\frac{\alpha^{n}}{\sqrt{n!}}\left|n\right\rangle .$
In terms of x-p coordinates, $\alpha=(1/2x_{0})(x+ip/M\omega)$.

The remaining higher order terms in the time-evolution operator originate
from the non-commutative property of the interaction Hamiltonian at
a given time with itself at different times. This can be understood
by considering the displacement operators, which do not commute with
one another but rather follow the commutation rule $\hat{D}(\alpha)\hat{D}(\beta)=\hat{D}(\alpha+\beta)e^{iIm(\alpha\beta^{*})}$.
Therefore the complete time-evolution operator can be constructed
by integrating over infinitesimal displacements in time:\begin{equation}
\hat{U}(t)=e^{i\Phi(t)}\hat{D}(\alpha(t)),\label{eq: evolution operator}\end{equation}
with the geometric phase accumulated over the entire path from time
$0$ to $t$ expressed as \begin{equation}
\Phi(t)=Im(\int_{0}^{t}\alpha(t')^{*}d\alpha(t')).\label{eq: geometric phase}\end{equation}
For a near-resonant driving force with detuning $\delta$ (Equation
\ref{eq: Interaction picture 2}), the initial motional state moves
in a circular trajectory of radius $F/(2\hbar\delta)$ with periodicity
$T=2\pi/\delta$ in the rotating frame of harmonic motion, following
the path (from Equation \ref{eq: alpha general})\begin{equation}
\alpha(t)=\frac{Fx_{0}}{2\hbar\delta}\left(1-e^{i\delta t}\right).\label{eq: alpha}\end{equation}
In one period of evolution under this force, the motional state returns
to its original phase space coordinates, but acquires a geometric
phase of \begin{equation}
\Phi_{0}=\frac{\pi\left|Fx_{0}\right|^{2}}{2(\hbar\delta)^{2}}\label{eq: Phi one period}\end{equation}
 equivalent to the area enclosed by the trajectory (Figure \ref{cap:Displacement}).

For a single qubit experiencing a spin-dependent force, the interaction
Hamiltonian includes a dependence on the internal spin state of the
ion. Assuming the force couples to only one of the vibrational modes
and the other mode can be neglected under the rotating wave approximation,
the most general expression for the interaction Hamiltonian is\begin{equation}
\hat{H}_{I}=\sum_{m=\uparrow_{\mathbf{n}},\downarrow_{\mathbf{n}}}\left(\frac{F_{m}^{*}x_{0}}{2}\hat{a}e^{-i\delta t}+\frac{F_{m}x_{0}}{2}\hat{a}^{\dagger}e^{i\delta t}\right)\left|m\right\rangle \left\langle m\right|,\label{eq: spin-dependent force single}\end{equation}
where $m$ denotes the internal qubit state of the ion, and $\left|\uparrow_{\mathbf{n}}\right\rangle $
and $\left|\downarrow_{\mathbf{n}}\right\rangle $ are the eigenstates
of $\hat{\mbox{\boldmath$\sigma$}}\cdot\mathbf{n}$ associated with eigenvalues
$+1$ and $-1$ respectively. The interaction provides no coupling
between the two orthogonal spin eigenstates of $\hat{\mbox{\boldmath$\sigma$}}\cdot\mathbf{n}$,
but the motional state becomes entangled with the spin state as the
differential force pushes the motional states of the two spin components
in separate directions. At time $t=2n\pi/\delta$, where $n$ is an
integer, the two motional states overlap again, disentangling the
vibrational component of the wavefunction from the spin, but leaving
the spin component with a phase shift due to the difference in the
geometric phases of the paths. The interaction Hamiltonian can also
be written in terms of the $\hat{\mbox{\boldmath$\sigma$}}\cdot\mathbf{n}$
operator as follows:\begin{equation}
\hat{H}_{I}=\left(\frac{F_{+}^{*}x_{0}}{2}\hat{a}e^{-i\delta t}+\frac{F_{+}x_{0}}{2}\hat{a}^{\dagger}e^{i\delta t}\right)\hat{I}+\left(\frac{F_{-}^{*}x_{0}}{2}\hat{a}e^{-i\delta t}+\frac{F_{-}x_{0}}{2}\hat{a}^{\dagger}e^{i\delta t}\right)\hat{\mbox{\boldmath$\sigma$}}\cdot\mathbf{n},\label{eq: H=sigma.n}\end{equation}
where $\hat{I}$ is the identity operator, $F_{+}=(F_{\uparrow_{\mathbf{n}}}+F_{\uparrow_{\mathbf{n}}})/2$
and $F_{+}=(F_{\uparrow_{\mathbf{n}}}-F_{\uparrow_{\mathbf{n}}})/2$.

Now consider a spin-dependent force applied simultaneously to two
ions in the same trapping potential. The total force on the system
is now dependent on the spins of both ions. The interaction Hamiltonian
now becomes\begin{equation}
\hat{H}_{I}=\sum_{m1,m2=\uparrow_{\mathbf{n}},\downarrow_{\mathbf{n}}}\left(\frac{F_{m1,m2}(t)x_{0}}{2}\hat{a}^{\dagger}+\frac{F_{m1,m2}^{*}(t)x_{0}}{2}\hat{a}\right)\left|m1,m2\right\rangle \left\langle m1,m2\right|,\label{eq: spin-dependent force general}\end{equation}
where $m_{1}$, $m_{2}$ denotes the internal qubit state of ion 1
and ion 2 respectively, and $F_{m1,m2}=F_{m1}+F_{m2}$ is the total
force applied to the state $\left|m1,m2\right\rangle $. The geometric
phase of an enclosed loop is proportional to $\left|F_{m_{1},m_{2}}x_{0}\right|^{2}/\delta$
(Equation \ref{eq: Phi one period}), which can be calibrated so that
the nonlinearity results in a wavefunction whose spins are not factorizable,
thus creating entanglement between two ions.

The following sections will provide specific examples of entangling
gates using spin-dependent forces. While the fundamental concept is
the same in both instances, the experimental requirements and the
susceptibility to various sources of phase decoherence are distinct.
We will discuss these cases in detail and provide some solutions for
phase control of these gates.

\subsection{The $\hat{\sigma}_{z}$-gate}

As the name implies, the $\hat{\sigma}_{z}$-gate applies a differential
force on the eigenstates of the unperturbed Hamiltonian. The interaction
has no coupling between the two eigenstates of $\hat{\sigma}_{z}$,
conveniently avoiding the neccessity to produce large frequency shifts
in the Raman beams to bridge the hyperfine splitting. Instead, the
Raman beams couple to the vibrational states without driving a spin
flip. The two Raman beams form a beating wave with periodicity $2\pi c/(\omega_{\nu}-\delta)$
along the weakest trap dimension. Due to the AC Stark effect, this
wave form a moving periodic potential, exerting a near-resonant force
on the ion in the direction of travel. If the AC Stark effect has
different amplitude on the two qubit states, then the two states experience
different forces from the beating wave \cite{leibfried03}.

\begin{figure}
\begin{center}
\includegraphics[
  width=8cm,
  height=6cm,
  keepaspectratio,
  angle=-90]{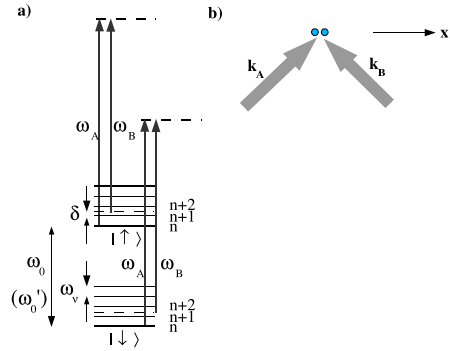}
\end{center}

\caption{A $\sigma_{z}$-dependent force is driven by electromagnetic fields
with two frequencies separated by $\omega_{\nu}+\delta$, as shown
in a). These fields couple the two qubit states to the excited states
with different coupling strengths (depending on polarization), producing
a differential AC Stark shift that oscillates at $\omega_{\nu}+\delta$.
The two fields must have a non-zero wave-vector difference $\Delta\mathbf{k}=\mathbf{k}_{B}-\mathbf{k}_{A}$
with a component in the $x$ direction.\label{cap: sigma_z}}
\end{figure}

The $\hat{\sigma}_{z}$ gate uses two non-copropagating beams with
frequency difference $\omega_{\nu}-\delta$, where $\omega_{\nu}$
is the frequency of vibration and $\delta$ is the detuning from the
vibration frequency. For this example, we will let the beams couple
to the stretch mode $\omega=\omega_{2}=\sqrt{3}\omega_{1}$, though
the same algebra can be carried out for the center-of-mass mode. (Stretch
mode is a better candidate since it exhibits lower levels of decoherence
from background electric fields \cite{turchette00a}.) We apply two
fields $E_{A}e^{i(\mathbf{k}_{A}\cdot\mathbf{x}-\omega_{A}t-\phi_{A})}\mbox{\boldmath$\epsilon$}_{A}+E_{B}e^{i(\mathbf{k}_{B}\cdot\mathbf{x}-\omega_{B}t-\phi_{B})}\mbox{\boldmath$\epsilon$}_{B}$
where the frequency difference $\omega_{B}-\omega_{A}=\omega_{2}-\delta$
is slightly detuned from the stretch mode frequency. The field couples
each of the spin states to the excited $P$ state, and is detuned
by a large frequency $\Delta$ (see Figure \ref{cap: sigma_z}). Using
the same RWA and adiabatic elimination of the excited state used to
obtain Equation \ref{AdElim_H}, the interaction Hamiltonian for a
single ion becomes\[
\hat{H}_{I}=\frac{\hbar}{4\Delta}\left\{ \left[\left|g_{\uparrow A}\right|^{2}+\left|g_{\uparrow B}\right|^{2}+\left(2g_{\uparrow A}^{*}g_{\uparrow B}e^{i(\Delta\mathbf{k}\cdot\hat{\mathbf{R}}-(\omega_{2}-\delta)t-\Delta\phi)}+h.c.\right)\right]\left|\uparrow\right\rangle \left\langle \uparrow\right|\right.\]
\begin{equation}
+\left.\left[\left|g_{\downarrow A}\right|^{2}+\left|g_{\downarrow B}\right|^{2}+\left(2g_{\downarrow A}^{*}g_{\downarrow B}e^{i(\Delta\mathbf{k}\cdot\hat{\mathbf{R}}-(\omega_{2}-\delta)t-\Delta\phi)}+h.c.\right)\right]\left|\downarrow\right\rangle \left\langle \downarrow\right|\right\} \label{eq: Stark shift}\end{equation}
where $g_{m,l}=\mbox{\boldmath$\mu$}_{m}\cdot\mbox{\boldmath$\epsilon$}_{l}E_{l}/2\hbar$
is the single photon Rabi frequencies associated with each field $l$
coupling qubit state $\left|m\right\rangle $ to excited level $\left|e\right\rangle $,
$\Delta\mathbf{k}=\mathbf{k}_{B}-\mathbf{k}_{A}$ is the wave vector
difference, and $\Delta\phi=\phi_{B}-\phi_{A}$ is the phase difference
between the driving fields. The first two terms in the expressions
in Equation \ref{eq: Stark shift} contribute to the average Stark
shift for each of the states, which can be canceled by carefully
choosing the polarizations $\mbox{\boldmath$\epsilon$}_{1}$ and $\mbox{\boldmath$\epsilon$}_{2}$
\cite{leibfried03}. The cross terms represent the variation in intensity
formed by the interference pattern that pushes the ion, and must have
a different magnitude and/or phase between the two qubit states to
create a differential force. In the Lamb-Dicke limit, and assume the
detuning $\Delta$ is approximately the same for both spin states
($\Delta\gg\omega_{0}$), the interaction Hamiltonian for two ions
can be written as \begin{eqnarray}
\fl \hat{H}_{I}=\frac{\hbar}{2}\sum_{i=1,2}\sum_{m_{i}=\uparrow,\downarrow}\eta_{2}\Omega_{m_{i}}D'_{n_{2},n_{2}'}\left(\hat{a}_{2}e^{-i(\delta t-\phi_{i})}+\hat{a}_{2}^{\dagger}e^{i(\delta t-\phi_{i})}\right)\left|m_{i}\right\rangle \left\langle m_{i}\right|\nonumber\\
\lo
=\sum_{m_{1},m_{2}=\uparrow,\downarrow}\left(\frac{F_{m_{1},m_{2}}^{*}q_{2}}{2}\hat{a}_{2}e^{-i\delta t}+\frac{F_{m_{1},m_{2}}q_{2}}{2}\hat{a}_{2}^{\dagger}e^{i\delta t}\right)\left|m_{1}m_{2}\right\rangle \left\langle m_{1}m_{2}\right|, \label{eq: z-gate interaction}
\end{eqnarray} where $F_{m_{1},m_{2}}q_{2}=(\hbar\eta_{2}D'_{n_{2,}n_{2}}/\Delta)(g_{m_{1},A}^{*}g_{m_{1},B}e^{i\phi_{1}}-g_{m_{2},A}^{*}g_{m_{2},B}e^{i\phi_{2}})$,
and $\phi_{i}=\Delta kX_{0,i}-\Delta\phi$.

\begin{figure}
\begin{center}
\includegraphics[
  width=8cm,
  keepaspectratio]{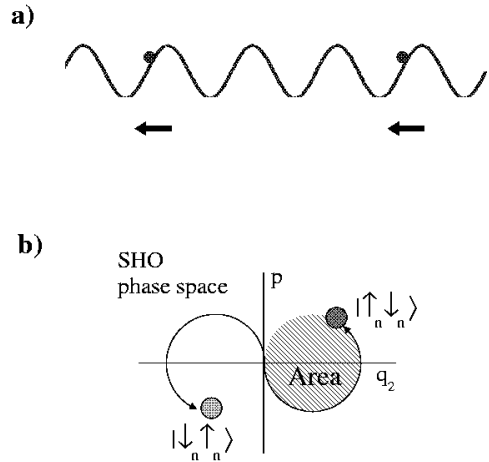}
\end{center}

\caption{The spacing of the two ions determines the relative phase of the
optical field experienced by each ion. a) The ions drawn are spaced
by an integer multiple of the optical wavelength, creating an equal
force on the two ions given the same internal state.  In this scenario,
b) spin states with opposite parity can excite the stretch mode, and
the geometric phase acquired by each state is proportional to the
area covered by the trajectory in phase space ($\pi/2$ for the phase
gate shown in Equation \ref{eq: geometric phase gate}). The spin states
with the same parity remain at the origin (not shown here) and acquire
no geometric phase. \label{cap:Two ions}}
\end{figure}

The phase difference between the force applied to the two ions is
determined by the optical phase difference $\phi_{1}-\phi_{2}$, which
corresponds to the ion spacing at equilibrium. If the ions are spaced
by an integer multiple of the optical wavelength, i.e. $\Delta k(X_{0,1}-X_{0,2})=2n\pi$,
then they experience the same phase in the force, i.e. $\phi_{1}=\phi_{2}$
(see Figure \ref{cap:Two ions}). This is a convenient case since
the forces cancel when the two spins are aligned in the same direction,
and displacement occurs only when the spins are anti-aligned. A physical
explanation of this scenario is that the stretch mode can be excited
only when the two ions are pushed in different directions or with
different magnitudes. The fastest gate time possible for this scheme
is when the anti-aligned states acquire a $\pi/2$ phase shift in
time $T=2\pi/\delta$, or in other words, a round-trip geometric phase
$\Phi_{0}=\pi\left|F_{\uparrow,\downarrow}q_{2}\right|^{2}/2(\hbar\delta)^{2}=\pi/2$.
Under these conditions, and assuming all the average Stark shifts
have been accounted for, the gate performs the operation \cite{leibfried03,Milburn00}:\begin{equation}
\begin{array}{rrr}
\left|\uparrow\uparrow\right\rangle  & \rightarrow & \left|\uparrow\uparrow\right\rangle \\
\left|\uparrow\downarrow\right\rangle  & \rightarrow & i\left|\uparrow\downarrow\right\rangle \\
\left|\downarrow\uparrow\right\rangle  & \rightarrow & i\left|\downarrow\uparrow\right\rangle \\
\left|\downarrow\downarrow\right\rangle  & \rightarrow & \left|\downarrow\downarrow\right\rangle \end{array}.\label{eq: geometric phase gate}\end{equation}
 With a phase shift of $-\pi/2$ on both qubits, the final state is
equivalent to the result from a standard phase gate in Equation \ref{eq: phase gate}.

Note that the end result is completely independent of the optical
phase of the drive field. The optical phase $\phi_{i}$ is absorbed
in the term $F_{m_{1},m_{2}}$, translating to a phase shift in $\alpha$
that defines the coherent state. Since the acquired geometric phase
depends only on the area enclosed by the trajectory, the phase of
the resulting state has no correlation to the phase of $F_{m_{1},m_{2}}$.
While the optical phase still needs to be coherent during a gate,
variations in phase between gates are acceptable since they have no
impact on the outcome.

Ideally, if the $\hat{\sigma}_{z}$ gate can be performed on magnetic
field insensitive states, the gate would be extremely phase stable.
Unfortunately, in the limit where the detuning from the excited state
is large compared to the hyperfine splitting, magnetic field insensitive
states have no differential Stark shift (see Appendix A). Therefore,
this gate is susceptible to decoherence from magnetic field fluctuations
if performed on magnetic field sensitive states, or if performed on
magnetic field insensitive states in the limit where detuning from
excited state is comparable to the hyperfine splitting, by spontaneous
emission from the excited state.

\subsection{$\hat{\sigma}_{\phi}$-gate}

We call the gate scheme proposed by M$\textnormal{\o}$lmer and S$\textnormal{\o}$rensen
\cite{molmer99,sorensen99} a $\hat{\sigma}_{\phi}$ gate because
the interaction is analogous to the $\hat{\sigma}_{z}$ gate operating
in a rotated basis. Although the original treatment describes the
interaction in a four-level ladder system, here we describe it in
terms of spins and displaced motional states as in section 2.1. The
M$\textnormal{\o}$lmer-S$\textnormal{\o}$rensen gate employs simultaneous
addressing of both ions with bichromatic fields, one detuned from
the red sideband of a vibrational mode by frequency $\delta$ and
the other from the blue sideband by $-\delta$. The two sidebands
have equal strength $\eta_{\nu}\Omega/2$ in the Lamb-Dicke Limit,
and once again we assume the force couples only to the stretch mode.
The interaction Hamiltonian is simply the sum of the red sideband
plus the blue sideband (from Equations \ref{rsb} and \ref{bsb})
with a detuning $\delta$:
\begin{eqnarray}
\fl \hat{H}_{I}=-\frac{\hbar}{2}\sum_{i=1,2}\eta_{2}\Omega_{i}D'_{n_{2},n'_{2}}(e^{i(\Delta k_{r}X_{0,i}-\Delta\phi_{r})}\hat{\sigma}_{+}^{(i)}\hat{a}_{2}e^{-i\delta t}+e^{i(\Delta k_{b}X_{0,i}-\Delta\phi_{b})}\hat{\sigma}_{+}^{(i)}\hat{a}_{2}^{\dagger}e^{i\delta t}+h.c.),\label{eq: sigma phi}\end{eqnarray}
where $\eta_{2}\Omega_{0}D'_{n_{2},n'_{2}}$ is the sideband Rabi
frequency, $\Delta k_{r}$ and $\Delta k_{b}$ are the wave-vector
difference for the red and blue sidebands, $X_{0,i}$ indicates the
equilibrium position of the $i$-th ion along the x-axis, and $\Delta\phi_{r}$
and $\Delta\phi_{b}$ are the phases of the red and blue sidebands
respectively. We can simplify this expression to:
\begin{eqnarray}
\fl \hat{H}_{I}= \sum_{i=1,2}\frac{F_{i}q_{2}}{2}\hat{\sigma}_{\phi_{S,i}}(e^{i\phi_{M,i}}\hat{a}_{2}e^{i\delta t}+e^{-i\phi_{M,i}}\hat{a}_{2}^{\dagger}e^{-i\delta t})\nonumber\\ \fl \indent =\sum_{m_{1}=\uparrow_{\phi_{S,1}},\downarrow_{\phi_{S,1}}}\sum_{m_{2}=\uparrow_{\phi_{S,2}},\downarrow_{\phi_{S,2}}}\left(\frac{F_{m_{1},m_{2}}^{*}q_{2}}{2}\hat{a}_{2}e^{i\delta t}+\frac{F_{m_{1}m_{2}}q_{2}}{2}\hat{a}_{2}^{\dagger}e^{-i\delta t}\right)\left|m_{1}m_{2}\right\rangle \left\langle m_{1}m_{2}\right|,\nonumber\\ \label{interaction Hamiltonian}
\end{eqnarray}
where \begin{equation}
\hat{\sigma}_{\phi_{S,i}}^{(i)}=\hat{\mbox{\boldmath$\sigma$}}_{i}\cdot\left[\cos(\phi_{S,i})\mathbf{x}+\sin(\phi_{S,i})\mathbf{y}\right]=\hat{\sigma}_{+}^{(i)}e^{-i\phi_{S,i}}+\hat{\sigma}_{-}^{(i)}e^{i\phi_{S,i}}.\label{eq: definition of sigma_phi}\end{equation}
Here $F_{i}=\hbar\eta_{2}\Omega_{i}D'_{n_{2},n'_{2}}/q_{2}$ is the
differential force on the $i$-th ion, $\phi_{S,i}=-(\Delta k_{r}X_{0,i}-\Delta\phi_{r}+\Delta k_{b}X_{0,i}-\Delta\phi_{b})/2$
is the spin phase of the $i$-th ion, $\phi_{M,i}=(\Delta k_{r}X_{0,i}-\Delta\phi_{r}-\Delta k_{b}X_{0,i}+\Delta\phi_{b})/2$
is the phase of the force on the $i$-th ion, and $F_{m_{1},m_{2}}=\pm F_{1}e^{-i\phi_{M,1}}\pm F_{2}e^{-i\phi_{M,2}}$
where $\pm F_{i}$ corresponds to the force on the spin state $m_{i}=\uparrow_{\mathbf{\phi}_{S,i}},\downarrow_{\mathbf{\phi}_{S,i}}$
respectively on the $i$-th ion. As in the $\hat{\sigma}_{z}$ gate,
we set the phase of the force acting on the two ions to be opposite,
i.e. $F_{1}e^{i\phi_{M,1}}=-F_{2}e^{i\phi_{M,2}}$, and choose $\delta$
and $F$ such that the round-trip geometric phase satisfies the condition
$\Phi_{0}=2\pi\left|F_{1}q_{2}\right|^{2}/(\hbar\delta)^{2}=\pi/2$.
Then the final state of the gate is equivalent to the final state
in Equation \ref{eq: geometric phase gate}, except $\left|\uparrow_{i}\right\rangle $
and $\left|\downarrow_{i}\right\rangle $, the eigenstates of $\hat{\sigma}_{z}^{(i)}$,
are replaced by $\left|\uparrow_{\mathbf{\phi}_{S,i}}\right\rangle $
and $\left|\downarrow_{\mathbf{\phi}_{S,i}}\right\rangle $, the eigenstates
of $\hat{\sigma}_{\mathbf{\phi}_{S,i}}^{(i)}$. This gate written
in the $\hat{\sigma}_{z}$ basis produces the following truth table:\begin{equation}
\begin{array}{ccl}
\left|\uparrow\uparrow\right\rangle  & \rightarrow & \frac{1}{\sqrt{2}}\left\{ \left|\uparrow\uparrow\right\rangle -ie^{i(\phi_{s1}+\phi_{s2})}\left|\downarrow\downarrow\right\rangle \right\} \\
\left|\uparrow\downarrow\right\rangle  & \rightarrow & \frac{1}{\sqrt{2}}\left\{ \left|\uparrow\downarrow\right\rangle -i\left|\downarrow\uparrow\right\rangle \right\} \\
\left|\downarrow\uparrow\right\rangle  & \rightarrow & \frac{1}{\sqrt{2}}\left\{ \left|\downarrow\uparrow\right\rangle -i\left|\uparrow\downarrow\right\rangle \right\} \\
\left|\downarrow\downarrow\right\rangle  & \rightarrow & \frac{1}{\sqrt{2}}\left\{ \left|\downarrow\downarrow\right\rangle -ie^{-i(\phi_{s1}+\phi_{s2})}\left|\uparrow\uparrow\right\rangle \right\} \end{array}\label{eq: sigma phi truth table}\end{equation}
Note that after the gate, only the spin phase remains, while the motion
phase has no effect on the final state. As in the $\hat{\sigma}_{z}$
gate, drifts in the motion phase between gates is acceptable. However,
the spin phase must be maintained between gates, or alternatively,
an equivalent entangling gate with the dependence on the spin phase
removed can be formed using a combination of $\hat{\sigma}_{\phi}$
gate and other quantum operations.

\begin{figure}
\begin{center}
\includegraphics[
  width=8cm,
  height=8cm,
  keepaspectratio]{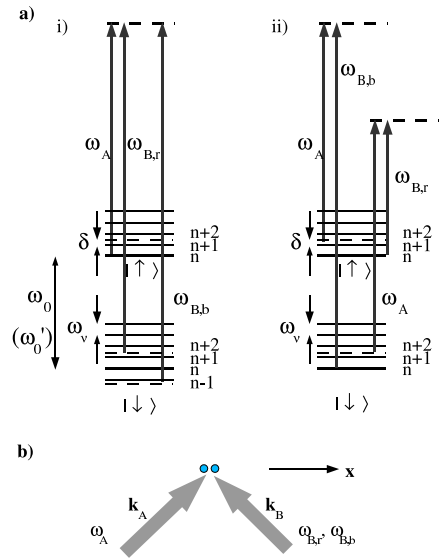}
\end{center}

\caption{A $\hat{\sigma}_{\phi}$-dependent force is driven by electromagnetic
fields with at least three optical frequencies as shown in a). Two
frequencies separated by $\omega_{0}'-\omega_{\nu}-\delta$ drive
a detuned red sideband and a third frequency differs from one of them
by $\omega_{0}'+\omega_{\nu}+\delta$ to drive a detuned blue sideband.
i) and ii) are two examples of possible frequency configurations.
Some of the fields can have overlapping wave-vectors, but any pair
of frequencies that drives a sideband must have a non-zero wave-vector
difference with a component in the $x$ direction. \label{cap: sigma phi}}
\end{figure}

An analysis of noise sources for the spin phase requires careful consideration
of the physical experimental setup in the laboratory. To drive the
red sideband and the blue sideband transitions simultaneously, a minimum
of three optical frequencies are required, assuming one frequency
can be used for both sideband couplings (see Figure \ref{cap: sigma phi}).
Since each of the two fields driving a sideband must have a non-zero
wave-vector difference $\Delta\mathbf{k}$, the optical beams can
be set up such that two fields at different frequencies share the
same wave-vector $\mathbf{k}_{B}$, and each individual field combined
with a third field having a different wave-vector $\mathbf{k}_{A}$
drives a sideband transition. In other words, if the field along $\mathbf{k}_{A}$
has frequency $\omega_{A}$, then the field propagating along $\mathbf{k}_{B}$
needs both a frequency component $\omega_{A}\pm(\omega_{0}'-\omega_{2}-\delta)$
to drive a detuned red sideband, and a frequency component $\omega_{A}\pm(\omega_{0}'+\omega_{2}+\delta)$
to drive a detuned blue sideband. The choice of the positive or negative
frequency differences between the fields determines the sign of $\Delta k_{r}$
and $\Delta k_{b}$, and determines the gate's susceptibility to the
phase stability between the two wave-vectors.

\subsubsection{Phase sensitive geometry}

\begin{figure}
\begin{center}
\includegraphics[%
  bb=0in 2in 8in 9in,
  clip,
  width=8cm,
  keepaspectratio]{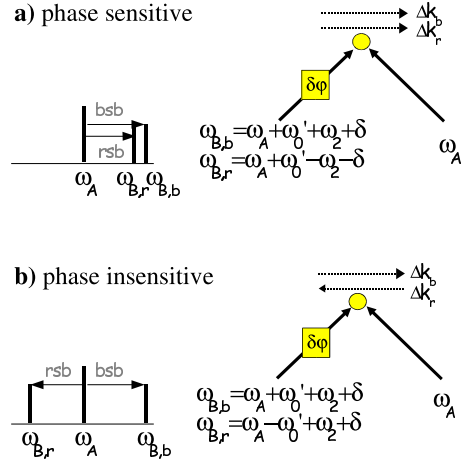}
\end{center}

\caption{Possible beam configurations for the $\sigma_{\phi}$ gate. a) Phase
sensitive configuration. The wave-vector difference for pairs of frequencies
driving the red sideband and the blue sideband travel in the same
direction $\Delta\mathbf{k}_{r}=\Delta\mathbf{k}_{b}$, using the
frequency configuration shown in Figure \ref{cap: sigma phi} a) i)
). A phase shift $\delta\phi$ in one beam path results in a phase
shift in the spin of the entangled state. b) Phase insensitive configuration.
The wave-vector difference for pairs of frequencies driving the red
sideband and the blue sideband travel in the opposite direction $\Delta\mathbf{k}_{r}=\Delta\mathbf{k}_{b}$,
using the frequency configuration shown in Figure \ref{cap: sigma phi}
a) ii). A phase shift $\delta\phi$ in one beam path results in no
net phase shift in the spin of the entangled state. \label{cap: beam configurations}}
\end{figure}

The first scenario involves frequencies of both fields along $\mathbf{k}_{B}$
being higher (or lower) than $\omega_{A}$, then the wave-vector difference
$\Delta\mathbf{k}$ for both the red and the blue sideband propagates
in the same direction (Figure \ref{cap: beam configurations}a). For
example, let the field along $\mathbf{k}_{B}$ include both $\omega_{A}+\omega_{0}'-\omega_{2}-\delta$
and $\omega_{A}+\omega_{0}'+\omega_{2}+\delta$ frequency components.
Then the wave-vector difference $\Delta\mathbf{k}_{r}=\mathbf{k}_{B}-\mathbf{k}_{A}=\Delta\mathbf{k}_{b}$
for the red sideband and the blue sideband point in the same direction.
Instability in the relative beam paths results in an equal phase shift
in the sideband transitions, i.e. $\delta\phi_{r}=\delta\phi_{b}=\delta\phi$.
This results in a net shift in the spin phase by $\delta\phi_{S,i}=\delta\phi$.
This is not a desirable situation since the outcome of the gate is
sensitive to changes in the beam path length difference on the scale
of an optical wavelength.

However, we note that the spin phase shift is exactly the same as
the phase shift on the non-copropagating carrier transition, i.e.
when the field propagating along $\mathbf{k}_{B}$ has frequency $\omega_{A}+\omega_{0}'$.
Therefore it is possible to construct a phase gate using the following
Ramsey experiment: 1) Perform a $\pi/2$ rotation on both ion with
phase shift $\delta\phi_{S,i}=\delta\phi$ using the non-copropagating
transition; 2) Perform the $\hat{\sigma}_{\phi}$ gate using the frequencies
listed above; 3) Perform a $-\pi/2$ rotation on both ions with phase
shift $\delta\phi_{S,i}=\delta\phi$ using the non-copropagating transition.
This rotation from $\mathbf{z}$ to $\mathbf{\phi}$ before the $\hat{\sigma}_{\phi}$
gate and the subsequent rotation back to $\mathbf{z}$ after the $\hat{\sigma_{\phi}}$
gate effectively removes the dependence on the spin phase $\phi$
as long as the spin phase is constant during the Ramsey experiment.
The final state becomes identical to Equation \ref{eq: geometric phase gate}
and has no residual dependence on $\delta\phi$. In addition, this
scheme is also insensitive to ion spacing since the phase of the push
force is always zero in the basis defined by $\phi_{S,i}$.

\subsubsection{Phase insensitive geometry}

Another scenario is to select the frequencies along $\mathbf{k}_{B}$
to straddle the frequency along $\mathbf{k}_{A}$. Then the wave-vector
difference for the red and the blue sideband propagates in opposite
directions. For example, let the field along $\mathbf{k}_{B}$ include
both $\omega_{A}-(\omega_{0}'-\omega_{2}-\delta)$ and $\omega_{A}+\omega_{0}'+\omega_{2}+\delta$
frequency components (Figure \ref{cap: beam configurations}b). Then
the wave-vector difference for the red sideband $\Delta\mathbf{k}_{r}=-\mathbf{k}_{B}+\mathbf{k}_{A}=-\Delta\mathbf{k}_{b}$
is in the opposite direction as the blue sideband. Instability in
the relative beam paths results in an opposite phase shift in the
sidebands, i.e. $-\delta\phi_{r}=\delta\phi_{b}=\delta\phi$. This
results in a net zero change in the spin phase $\delta\phi_{S,i}=0$,
removing any spin phase dependence on $\delta\phi$ from the gate.
Hence this configuration is termed {}``phase insensitive''.

However, the motion phase in this setup acquires a dependence on the
phase shift $\delta\phi$. Therefore the phase of the force on each
ion should be calibrated to be the same by setting the ion spacing
(using the trap frequency as a tuning parameter) equal to $X_{0,1}-X_{0,2}=2n\pi/\Delta k$,
where $n$ in an integer. While it is possible to produce similar
entanglement operations with other values of ion spacing, the gate
speed will be slower for the same intensity from the laser, and the
output will appear different than the expression in Equation \ref{eq: sigma phi truth table}.

Similar to the other configuration, it is possible to construct a
phase gate with the transformation in Equation  \ref{eq: geometric phase gate},
using an analogous Ramsey experiment involving single qubit rotations
in phase with the gate: 1) Perform a $\pi/2$ rotation on both ion
with phase shift $\delta\phi_{S,i}=0$ using either a calibrated and
phase locked microwave source or a copropagating Raman transition;
2) Perform the $\sigma_{\phi}$ gate using the frequencies listed
here; 3) Perform a $-\pi/2$ rotation on both ions with phase shift
$\delta\phi_{S,i}=0$.

\section{$\sigma_{z}$-gate with fast pulses}

The $\sigma_{z}$-gate can also be achieved by applying spin-dependent
momentum kicks on the ions with fast laser pulses \cite{Garcia-Ripoll03,Duan04}.
For gates in the resolved-sideband limit discussed in section
2, the ion is assumed to be confined within the Lamb-Dicke limit,
where the spread in the position of the ions from their equilibrium
positions is much smaller than the optical wave length. Outside of
this limit, the effective Rabi frequency fluctuations leads to significant
gate errors. For gates using fast laser pulses \cite{Garcia-Ripoll03,Duan04},
the impulsive spin-dependent force from the traveling wave has an
almost uniform intensity distribution around the ion's position, and this
fluctuation of the effective Rabi frequency can be safely neglected.
These pulsed gates can therefore faithfully operate outside of the
Lamb-Dicke limit. In this section, we want to show that non-trivial
phase errors can arise when the ions are outside the Lamb-Dicke limit,
and suggest a method to cancel these errors by carefully selecting
the direction and timing of momentum transfer, a technique reminiscent
of the phase cancellation effect in the phase insensitive $\sigma_{\phi}$
gate configuration discussed in section 2.3.2.

The central component of the fast $\sigma_{z}$ gate in the context
of ground state hyperfine qubits is a set of fast resonant laser pulse
pairs that exclusively couple one of the two qubit states (here taken
to be $\left|\downarrow\right\rangle $) of each ion to the excited
state $\left|e\right\rangle $. The coupling Hamiltonian follows from
from Equation 2 with $\Delta=0$:\begin{equation}
\hat{H}_{I}=\hbar\sum_{i=1,2}-\left(\frac{g(t)}{2}e^{ik\cdot X_{0,i}-i\phi}e^{-i\eta_{1}(\hat{a}_{1}+\hat{a}_{1}^{\dagger})\mp i\eta_{2}(\hat{a}_{2}+\hat{a}_{2}^{\dagger})}\left|e_{i}\right\rangle \left\langle \downarrow\right|+h.c.\right)\label{eq: H_fast}\end{equation}
where $g(t)$ is the resonant Rabi frequency of the transition for
each ion and as before, the plus (minus) sign refers to ion $i=1$
($i=2$). The pulse pairs are set to drive successive $\pi$-pulses
($\int_{0}^{\tau}g(t)dt=\pi$) from $\left|\downarrow_{i}\right\rangle \rightarrow\left|e_{i}\right\rangle \rightarrow\left|\downarrow_{i}\right\rangle $
on the electronic transitions of each ion, with the pulse duration
$\tau$ taken to be much shorter than the radiative lifetime of $\left|e\right\rangle $
as well as the trap period $2\pi/\omega_{\nu}$.

When these two successive fast pulses have non-copropagating wave-vectors
$\mathbf{k}_{A}$ and $\mathbf{k}_{B}$, and both $\pi$-pulses are
completed in a time much shorter than the lifetime of state $\left|e_{i}\right\rangle $,
the result is the following evolution for two ions \cite{Blatt96,Garcia-Ripoll03}:\begin{equation}
\begin{array}{lll}
\uparrow_{1}\uparrow_{2}\left|\alpha\right\rangle _{1}\left|\alpha\right\rangle _{2} & \rightarrow & \uparrow_{1}\uparrow_{2}\left|\alpha\right\rangle _{1}\left|\alpha\right\rangle _{2}\\
\uparrow_{1}\downarrow_{2}\left|\alpha\right\rangle _{1}\left|\alpha\right\rangle _{2} & \rightarrow & \uparrow_{1}\downarrow_{2}\left|\alpha+i\eta_{1}\right\rangle _{1}\left|\alpha-i\eta_{2}\right\rangle _{2}\\
\downarrow_{1}\uparrow_{2}\left|\alpha\right\rangle _{1}\left|\alpha\right\rangle _{2} & \rightarrow & \downarrow_{1}\uparrow_{2}\left|\alpha+i\eta_{1}\right\rangle _{1}\left|\alpha+i\eta_{2}\right\rangle _{2}\\
\downarrow_{1}\downarrow_{2}\left|\alpha\right\rangle _{1}\left|\alpha\right\rangle _{2} & \rightarrow & \downarrow_{1}\downarrow_{2}\left|\alpha+2i\eta_{1}\right\rangle _{1}\left|\alpha\right\rangle _{2}\end{array}.\label{eq: H_fast}\end{equation}
In this expression, $\left|\alpha_{1}\right\rangle $ and $\left|\alpha_{2}\right\rangle $
are initial coherent states of the two modes of motion, and $\eta_{\nu}$
are the Lamb-Dicke parameters of the two modes associated with the
wave-vector difference $\Delta\mathbf{k}_{j}\equiv\mathbf{k}_{A}-\mathbf{k}_{B}$,
exactly as defined in section 1.1.

In the fast $\sigma_{z}$ gate, a series of pulse pairs is applied
to the ions so that the motional states of both modes of motion simultaneously
return to the same position in phase space regardless of the state
of the two qubits. When these fast pulses are interspersed with periods
of free evolution of the two modes of harmonic motion, the result
can be a $\sigma_{z}$ phase gate for appropriate choices of pulse
timing \cite{Garcia-Ripoll03,Duan04}. This fast gate works independent
of the motional state and outside of the Lamb-Dicke limit, as long
as the motion remains harmonic.

However, outside of the Lamb-Dicke regime, we find that this gate
can be sensitive to changes in the phase of the optical fields due
to the change in the position of the ions at different times. In order
to see this effect, we note that this fast $\sigma_{z}$-gate involves
a spin-dependent force on the ion from absorption of a photon from
a laser pulse traveling in the $\mathbf{k}_{A}$ direction and emission
of another photon to a pulse propagating in a different $\mathbf{k}_{B}$
direction. We can lock the relative phase of these two propagating
laser beams so that their phase difference is set to zero at the ion's
equilibrium position. Then, if during the above impulsive kicks, the
ion is at a position $\mathbf{r}$ from its equilibrium site, it will
acquire a net spin-dependent phase factor of $e^{i\Delta\mathbf{k}\cdot\mathbf{r}}$
from the absorption and the emission of the photon. This phase factor
from each spin-dependent kick is non-negligible if the ion is outside
of the Lamb-Dicke limit. For a complete gate operation with a series
of laser kicks, with the spin-dependent phase factor for the $j$-th
kick is denoted by $e^{i\Delta\mathbf{k}_{j}\cdot\mathbf{r}_{j}}$,
the total phase factor after $N$ kicks is given by $e^{i\varphi_{t}}$
with $\varphi_{t}=\sum_{j=1}^{N}\left(\Delta\mathbf{k}_{j}\cdot\mathbf{r}_{j}\right)$.
If the gate speed is comparable or slower than the local ion oscillation
frequency, the ion's position $\mathbf{r}_{j}$ at different laser
kicks changes depending on the initial momentum and position are become
almost uncorrelated. Therefore, the above effect contributes a random
phase to the spin, which is a source of the gate infidelity.

To eliminate this random phase effect when the ion is outside the
Lamb-Dicke limit, one needs to require the gate speed to be significantly
faster than the local ion oscillation frequency. In that case, the
ion's position at different laser kicks are almost the same although
they are still unknown. For any two positions $\mathbf{r}_{j}$ and
$\mathbf{r}_{k}$ during the $j$-th and $k$-th kick respectively,
the difference between them can be bounded as $\left|\mathbf{r}_{k}-\mathbf{r}_{j}\right|\lesssim vT_{g}$,
where $v$ is the ion's typical speed and $T_{g}$ denotes the gate
time. Due to this position correlation and the fact that the total
momentum kicks $\sum_{j=1}^{N}\Delta\mathbf{k}_{j}=0$ for the fast
gate, we conclude that the random phase $\varphi_{t}$ is bounded
by $\varphi_{t}\lesssim\left|\Delta\mathbf{k}_{j}\right|vT_{g}$,
and the gate infidelity $\delta F\equiv1-F$ from this random phase
scales as $\left(\left|\Delta\mathbf{k}\right|vT_{g}\right)^{2}$.
The scaling of $\delta F$ can be further improved to $\left(\left|\Delta\mathbf{k}\right|vT_{g}\right)^{2n}$
if we use a more involved sequence of the kicking forces with $n$
basic cycles. The gate time must be short enough to make the scaling
parameter $\left|\Delta\mathbf{k}\right|vT_{g}<1$. Under that condition,
the gate infidelity can then be reduced rapidly to zero with an appropriate
pulse sequence even if the ion is outside of the Lamb-Dicke limit
\cite{Duan04}.

\section*{Conclusion}

Most quantum logic gate schemes for trapped ions operate through interactions
with optical electromagnetic fields. Some schemes, such as the $\sigma_{\phi}$
gate and the fast $\sigma_{z}$ gate, have a phase dependence on the
phase of the optical driving field, which can become a major source
of decoherence if uncontrolled. We have shown here methods to remove
this phase dependence for these two entangling gates by choosing appropriate
wave-vector orientations and pulse timings that naturally cancel the
phase factor $e^{i\Delta\mathbf{k}\cdot\mathbf{r}}$ upon the completion
of the gate. Furthermore, the sideband resolved $\sigma_{\phi}$ gate
can operate on magnetic field insensitive qubit states, removing an
unavoidable vulnerability of the $\sigma_{z}$ gate. These techniques
eliminates the random phase from the optical driving field while maintaining
phase coherence at the rf or microwave atomic frequencies, allowing
long gate sequences to be performed over the timescales beyond the coherence time of the
optical fields.

\section*{Acknowledgments}

We would like to thank Winfried Hensinger for helpful discussions.
This work is supported by the U.S. National Security Agency and Advanced
Research and Development Activity under Army Research Office contract
DAAD19-01-10667 and the National Science Foundation Information Technology
Research Program.

\appendix
\section{Magnetic field insensitivity}

In this appendix we will show that magnetic field insensitive states
have no differential Stark shift in the limit where the detuning from
the excited state is much larger than the hyperfine splitting, i.e.
$\Delta_{HF}/\Delta\rightarrow0$ \cite{Langer05}. To find the field insensitive states
for a system in the $S_{1/2}$ ground state with some nuclear spin
$I$, we write down the Hamiltonian for the hyperfine interaction
in the presence of a magnetic field $\mathbf{B}$:
\begin{equation}
\hat{H}=\mbox{\boldmath$\mu$}_{B}\cdot\mathbf{B}+A\hat{\mathbf{I}}\cdot\hat{\mathbf{J}}=g_{J}\mathbf{B}\cdot\hat{\mathbf{J}}+g_{I}\mathbf{B}\cdot\hat{\mathbf{I}}+A\hat{\mathbf{I}}\cdot\hat{\mathbf{J}},\label{eq: hyperfine_interaction}\end{equation}
 where $\hat{\mathbf{J}}$ is the total angular momentum of the electron,
$\hat{\mathbf{I}}$ is the nuclear spin, and $A\hat{\mathbf{I}}\cdot\hat{\mathbf{J}}$
is the contact term. $g_{I}$ and $g_{J}$ are the Lande g-factors
for the nucleus and the electron. The eigenstates of the Hamiltonian
are linear combinations of the $m_{F}$ states, and can be represented
as $\left|\Psi_{i}\right\rangle =a_{i}\left|g;m_{J}=\frac{1}{2},m_{I}=m_{F,i}-\frac{1}{2}\right\rangle +b_{i}\left|g;m_{J}=-\frac{1}{2},m_{I}=m_{F,i}+\frac{1}{2}\right\rangle $.
The coefficients $a$ and $b$ are functions of the magnetic field.
If two states $\left|\Psi_{1}\right\rangle $ and $\left|\Psi_{2}\right\rangle $
are magnetic field insensitive, then
\begin{equation}
\frac{\partial}{\partial B}\left(E_{1}-E_{2}\right)=0.\label{eq: partial B}\end{equation}
Applying Ehrenfest's theorem,
\begin{equation}
\fl \frac{\partial E_{i}}{\partial B}=\left\langle g_{J}J_{z}+g_{I}I_{z}\right\rangle =\left|a_{i}\right|^{2}\left[\frac{g_{J}}{2}+g_{I}(m_{F,i}-\frac{1}{2})\right]+\left|b_{i}\right|^{2}\left[\frac{g_{J}}{2}+g_{I}(m_{F,i}+\frac{1}{2})\right].\label{eq: dE/dB}\end{equation}
Normalization of the eigenstates and solving Equation \ref{eq: partial B}
gives the result $\left|a_{1}\right|^{2}=\left|a_{2}\right|^{2}+g_{I}\Delta m_{F}/(g_{J}-g_{I})$.
Since the dipole moment of the electron dominates the dipole moment
of the nucleus, i.e. $g_{I}/g_{J}\approx10^{-3}$, we can approximate
it as \[
\left|a_{1}\right|^{2}=\left|a_{2}\right|^{2}\]
\begin{equation}
\left|b_{1}\right|^{2}=\left|b_{2}\right|^{2}.\label{eq:coef_relation}\end{equation}

Now consider the Stark shift for each of these magnetic field insensitive
states. The AC Stark shift is given by\begin{equation}
\chi_{i}=\sum_{m_{J},m_{I}}\frac{\left\langle \Psi_{i}\right|E\cdot d\left|e,m_{J}m_{I}\right\rangle \left\langle e,m_{J}m_{I}\right|E\cdot d\left|\Psi_{i}\right\rangle }{\Delta-E_{1}+E_{i}},\label{eq: Stark shift A}\end{equation}
where $\left|e;m_{J},m_{I}\right\rangle $ is the excited state with
the corresponding z-component of the electron and nuclear spins. Since
the electric dipole only couples the orbital angular momentum of the
electron, $\Psi_{i}$ only couples to the states with the same $m_{I}$.
So the expression can be simplified to\begin{equation}
\fl \chi_{i}=\left|a_{i}\right|^{2}\sum_{m_{J}}\frac{\left|\left\langle g;m_{J}=\frac{1}{2}\right|E\cdot d\left|e,m_{J}\right\rangle \right|^{2}}{\Delta-E_{1}+E_{i}}+\left|b_{i}\right|^{2}\sum_{m_{J}}\frac{\left|\left\langle g;m_{J}=-\frac{1}{2}\right|E\cdot d\left|e,m_{J}\right\rangle \right|^{2}}{\Delta-E_{1}+E_{i}}.\label{eq: Stark shift A2}\end{equation}
If the energy difference between the two states $\Psi_{1}$ and $\Psi_{2}$
is small compared to $\Delta$, and applying the results from equation
\ref{eq:coef_relation}, then we find that $\chi_{1}=\chi_{2}$. So
we conclude that the energy shift due to Stark effect is the same
for any two magnetic field insensitive states.

\section{Driving stimulated Raman transitions using an Electro-optic
Modulator}

The fields driving stimulated Raman transition in ions are typically
generated from a single laser source with the multiple frequencies
generated by optical modulators. Acousto-optic modulators can produce
frequency shifts up to about 1 GHz, while electro-optic modulators
can modulate at upwards of 10 GHz or more. The electro-optic modulator
offers a solution for driving Raman transitions in ion species with
a large hyperfine splitting, but unlike acousto-optic modulators,
all frequencies in the modulated field have the same wave-vector, making it difficult to separate different frequency
components.

Electro-optic modulators control the birefringence of uniaxial crystals
with a lower frequency electric field, effectively modulating either
the phase or the polarization of the incident optical field, depending
on the orientation of the optical axes. Since the Raman coupling is
polarization-dependent, the polarization modulation is equivalent
to an amplitude modulation. For non-copropagating Raman beam geometry,
the modulated field is divided using a beam splitter and the two beams
recombine at the trap from different angles. If the difference between
any frequency from one beam and any frequency from the other beam
matches the energy splitting of two atomic and/or phonon levels, then
a transition can potentially be driven. However, the amplitude of each pair of frequencies
driving the transition can result in cancellations in the total transition
rate. Usually, amplitude modulation produces sidebands with the same
phase, resulting in the transition rates adding constructively. But
phase modulation produces a comb of sidebands, some having amplitudes
with opposite phases, which could result in a total transition rate
of zero. This problem can be remedied by setting the beam path length
difference between the two beams to certain values that produces a
non-zero total transition rate \cite{Lee03}. This effectively adds
a different phase to each sideband, resulting in a transition rate
proportional to the squared electric field \begin{equation}
\Gamma_{k}(\phi)\sim \sum_{n=-\infty}^{\infty}J_{n}(\phi)e^{in\theta}J_{n+k}(\phi)e^{i(n+k)\theta}=J_{k}\left(2\phi\sin\left(\theta\right)\right)\label{eq: Bessel_sum}\end{equation}
where $J_{n}(x)$ is the $n$-th order Bessel function.  Equation \ref{eq: Bessel_sum} describes the Raman transition rate involving the optical carrier and the $k$-th
frequency modulated sideband with modulation index $\phi$ with a phase
shift of $\theta=(\delta k\Delta x)mod(2\pi)$. Here $\delta k$ is
the wave-number associated with the modulation frequency and $\Delta x$
is the beam path length difference.

To avoid nulling the average intensity due to destructive interference
between the two fields at the ion, the field propagating along one
path can be frequency shifted slightly from the field in the other
path. The modulation frequency would then have to be adjusted to compensate
for this frequency shift so that pairs of frequencies matches the
energy difference between the two coupled levels. When the frequency
offset is accounted for, all the $\Delta\mathbf{k}$ vectors driving
the transition have the same sign, and the resulting transition is
exactly the same as if each of the two beams having only a single
frequency. To reverse the $\Delta\mathbf{k}$ vector, only the frequency
offset on one beam needs to be changed. For example, to drive a Raman
transition between two levels which have an energy splitting of $\hbar\omega_{transition}$,
the modulation frequency of the EOM can be set to $\omega_{EO}<\omega_{transition}$.
The modulated beam is split into two paths, with the beam in path
A frequency shifted by $\omega_{offset}$ and the beam in path B frequency
shifted by $\omega_{offset}+\omega_{transition}-\omega_{EO}$. The
two beams are recombined at the ion, with wave-vector $\mathbf{k}_{A}$
and $\mathbf{k}_{B}$ respective to beam path A and B. In this case
the wave-vector difference $\Delta\mathbf{k}$ for the Raman transition
is equal to $\mathbf{k}_{B}$$-\mathbf{k}_{A}$, since the beam in
path B has higher frequency. The beam in path B can also be frequency
shifted by $\omega_{offset}+\omega_{EO}-\omega_{transition}$ instead,
in which case the wave-vector difference $\Delta\mathbf{k}$ would
be $\mathbf{k}_{A}-\mathbf{k}_{B}$ since the beam in path A would
have the higher frequency.

The reversal of $\Delta\mathbf{k}$ is useful in the phase stable
configuration for a $\sigma_{\phi}$ gate (see section 2.3.2). To generate red and blue
sidebands with opposite wave-vector difference $\Delta\mathbf{k}_{r}=-\Delta\mathbf{k}_{b}$
, the frequency of the field along $\mathbf{k}_{B}$ can be shifted
by $\omega_{offset}+\omega_{0}'-\omega_{2}-\delta-\omega_{EO}$ to
generate the red sideband and by $\omega_{offset}-\omega_{0}'-\omega_{2}-\delta+\omega_{EO}$
to generate the blue sideband. These two frequencies can be arbitrarily
close to one another by tuning the modulation frequency of the EO
close to the qubit frequency splitting $\omega_{0}'$, which allows
both frequencies to be generated using a single frequency shifter
with a given bandwidth. However, if the modulation frequency of the
EO is exactly $\omega_{0}$, then each beam would simultaneously drive
a copropagating carrier transition (see section 1.2.1) that will interfere with the $\sigma_{\phi}$
operation. Therefore the modulation frequency should be tuned to approximately
but not exactly $\omega_{0}'$.

\section*{References}

\newpage

\noappendix


\begin{figure}
\begin{center}
\includegraphics[
  width=8cm,
  height=8cm,
  keepaspectratio]{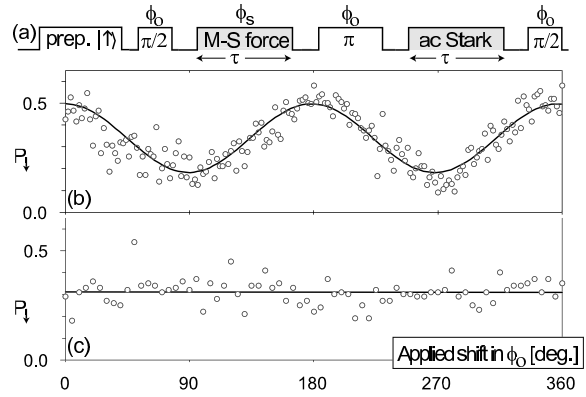}
\end{center}

\caption{Demonstration of the phase sensitivity of the M$\textnormal{\o}$lmer-S$\textnormal{\o}$rensen
($\sigma_{\phi}$) force applied to a single ion using different beam
configurations. (a) Pulse sequence of a photon-echo experiment for
testing optical phase sensitivity of the $\sigma_{\phi}$ force. The
echo pulses are carrier transitions driven by non-copropagating Raman
beams propagating along the same wave-vectors as the $\sigma_{\phi}$
force, and an applied phase shift $\phi_{0}$ is added using an acousto-optic
modulator that controls the timing of the pulse. The $\sigma_{\phi}$
force is applied for sufficient time such that the two motional states
corresponding to spin states $\left|\uparrow_{\phi}\right\rangle $
and $\left|\downarrow_{\phi}\right\rangle $ have very little overlap
at time $\tau$. A separate pulse in the other arm of the echo experiment
cancels the residual AC Stark shift induced by the field driving the
$\sigma_{\phi}$ force (the ratio $\omega_{0}/\Delta$ is significant
enough to produce a non-negligible differential Stark shift between
the two qubit states in this experiment) . (b) Probability of detecting
$\left|\downarrow\right\rangle $ vs. applied shift in the phase of
the echo pulses $\phi_{0}$ using the phase sensitive configuration
described in section 2.3.1 and Figure \ref{cap: beam configurations}a.
The fringe contrast shows coherence between the phase $\phi$ in the
$\sigma_{\phi}$ force and the phase $\phi_{0}$ in the Raman carrier
pulses (Probability should vary sinusoidally from 0 to 0.5 when there
is no decoherence). (c) Same plot using the phase insensitive configuration
described in section 2.3.2 and Figure \ref{cap: beam configurations}b.
This time there is no coherence between the phase-insensitive $\sigma_{\phi}$
force and the phase-sensitive non-copropagating Raman carrier pulses. }
\end{figure}

\begin{figure}
\begin{center}
\includegraphics[
  width=8cm,
  height=8cm,
  keepaspectratio,
  angle=-90]{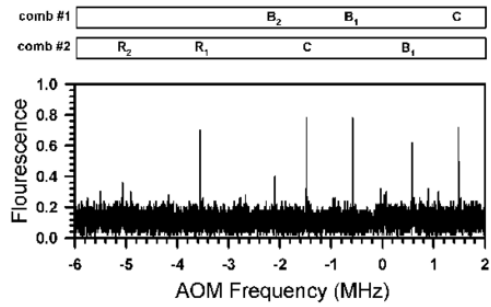}
\end{center}

\caption{Raman spectrum of two ions in the trap using a frequency comb generated
by an electro-optic modulator (modulation frequency $\omega_{EO}-\omega_{0}=1.5$MHz
in this case). The x-axis shows the frequency difference between the
fields along the two beam paths. The carrier transition appears at
$\pm1.5$MHz ($C$), with corresponding first center-of-mass blue
sideband transition at $\mp0.6$MHz ($B_{1}$) $(\omega_{1}/2\pi=2.1$MHz)
, first center-of-mass red sideband transition at $\pm3.6$MHz ($R_{1}$),
first stretch mode blue sideband at $\mp2.1$MHz ($B_{2}$) $(\omega_{2}/2\pi=3.6$MHz),
and first stretch mode red sideband at $\pm5.1$MHz ($R_{2}$). \label{cap: Raman spectrum}}
\end{figure}

\end{document}